\documentclass[structabstract]{aa}  
\usepackage{graphicx}
\usepackage{natbib}
\usepackage{txfonts}
\usepackage{color}
\begin{document}
   \title{The relation between chemical abundances and kinematics of the Galactic disc with RAVE}

   \author{
	C. Boeche\inst{1,2},  C. Chiappini\inst{2},
	I. Minchev\inst{2}, 
	M. Williams\inst{2}, M. Steinmetz\inst{2},
	S. Sharma\inst{3},
	G. Kordopatis\inst{4},
	J. Bland-Hawthorn\inst{3},
	O. Bienaym\'e\inst{5}, 
	B. K. Gibson\inst{6,7}, 
	G. Gilmore\inst{4},
	E. K. Grebel\inst{1}, 
	A. Helmi\inst{8},
	U. Munari\inst{9},
	J.F. Navarro\inst{10}, 
	Q. A. Parker\inst{11,12,13}, 
	W. Reid\inst{13},
	G. M. Seabroke\inst{14}, 
	A. Siebert\inst{4}, 
	A. Siviero\inst{15,2},
	F. G. Watson\inst{12},
	R. F. G. Wyse\inst{16},
	T. Zwitter\inst{17,18}
	}          

   \offprints{corrado@ari.uni-heidelberg.de}

   \institute{
Astronomisches Rechen-Institut, Zentrum f\"ur Astronomie der Universit\"at Heidelberg,
M\"onchhofstr. 12-14, 69120 Heidelberg, Germany
	\and
Leibniz-Institut f\"ur Astrophysik Potsdam (AIP), An der Sternwarte 16, 
14482 Potsdam, Germany
         \and
Sydney Institute for Astronomy, University of Sydney, NSW 2006, Australia
	\and
Institute of Astronomy, University of Cambridge, Madingley Road, Cambridge CB3 0HA, UK
	\and
Observatoire de Strasbourg, Universit\'e de Strasbourg, CNRS
11 rue de l'universit\'e, F-67000 Strasbourg, France
	\and
Jeremiah Horrocks Institute, University of Central Lancashire, Preston, PR1~2HE, UK
	\and
Monash Centre for Astrophysics, School of Mathematical Sciences, Monash University, Clayton, VIC, 3800, Australia
	\and
Kapteyn Astronomical Institute, University of Groningen, P.O. Box 800, 9700 AV Groningen,
The Netherlands
	\and
INAF Osservatorio Astronomico di Padova, Via dell'Osservatorio 8, Asiago I-36012, Italy
	\and
University of Victoria, P.O. Box 3055, Station CSC, Victoria, BC V8W 3P6, Canada
	\and
Department of Physics and Astronomy, Faculty of Sciences, Macquarie University, Sydney, NSW 2109, Australia
	\and
Anglo-Australian Observatory, P.O. Box 296, Epping, NSW 1710, Australia
	\and
Macquarie Research Centre for Astronomy, Astrophysics and Astrophotonics, Sydney, NSW 2109, Australia
	\and
Mullard Space Science Laboratory, University College London, Holmbury St Mary, Dorking, RH5 6NT, UK
	\and
Department of Physics and Astronomy, Padova University, Vicolo
 dell’Osservatorio 2, I-35122 Padova, Italy
	\and
Department of Physics and Astronomy, Johns Hopkins University, 3400 North
Charles Street, Baltimore, MD 21218, USA
	\and
Faculty of Mathematics and Physics, University of Ljubljana, Jadranska 19, SI-1000 Ljubljana, Slovenia
	\and
Center of Excellence SPACE-SI, Askerceva cesta 12, SI-1000 Ljubljana, Slovenia
}


 
  \abstract
   {}
  {We study the relations between stellar kinematics and chemical
abundances of a large sample of RAVE giants in search for selection criteria
needed for disentangling different Galactic stellar populations, such as thin
disc, thick disc and halo.  A direct comparison between the chemo-kinematic
relations obtained with our medium spectroscopic resolution data and those
obtained from a high-resolution sample is carried out with the aim of
testing the robustness of the RAVE data.}
   {We select a sample of 2167 giant stars with signal-to-noise per spectral
measurements above 75 from the RAVE chemical catalogue and follow the
analysis performed by Gratton and colleagues on 150 subdwarf stars
spectroscopically observed at high-resolution.  We then use a larger sample
of 9131 giants (with signal-to-noise above 60) to investigate the chemo-kinematical
characteristics of our stars by grouping them into nine subsamples with
common eccentricity ($e$) and maximum distance achieved above the Galactic
plane ($Z_{\rm max}$).}
 {The RAVE kinematical and chemical data proved to be reliable by
reproducing the results by Gratton et al.  obtained with high-resolution
spectroscopic data.  We successfully identified three stellar populations
which could be associated with the Galactic thin disc, a dissipative
component composed mostly of thick-disc stars, as well as a component
comprised of halo stars (presence of debris stars cannot be excluded). 
Our analysis, based on the $e$-$Z_{\rm max}$ plane combined with additional
orbital parameters and chemical information, provides an alternative way of
identifying different populations of stars.  In addition to extracting
canonical thick- and thin-disc samples, we find a group of stars in the
Galactic plane ( $Z_{\rm max} < $ 1~kpc and 0.4 $< e < $0.6), which show
homogeneous kinematics but differ in their chemical properties.  We
interpret this as a clear sign that some of these stars have experienced the
effects of heating and/or radial
migration, which have modified their original orbits.  The
accretion origin of such stars cannot be excluded.}
  {}

   \keywords{Galaxy: abundances -- Galaxy: evolution -- Galaxy: structure --
Galaxy: kinematics and dynamics
               }

    \authorrunning{C. Boeche et al.}

   \maketitle
%

\section{Introduction}

The chemical enrichment of the Universe is one of the main thrusts of modern
astrophysics, and the Milky Way (MW) can be seen as the Rosetta stone of this
evolution. In particular, Galactic archeology, i.e., the combined study of
kinematics and 
chemical composition of stars of different ages, has recently become one of the
cornerstones of research in galaxy formation. The main goal in this new field is to reconstruct 
the formation history of our Galaxy by analyzing its fossil chemical
records and kinematical information.  An important sub-product of this kind of
study is also to provide observational constraints to models of galaxy formation in general. 
While there has been considerable progress in the past years to reproduce
realistic disc galaxies in cosmologial simulations (e.g.  Guedes et al. 
\citealp{guedes11}, Brook et al.  \citealp{brook12a}, Stinson et al. 
\citealp{stinson}), the outcome heavily relies on the advanced feedback
models involving a considerable number of free parameters (Piontek \&
Steinmetz \citealp{piontek11}, Scannapieco et al.  \citealp{scannapieco11}) which
need to be calibrated using empirically determined relations.

Two main paths have been taken in the last years in the field of Galactic
archeology.  On the one hand high-resolution, high signal-to-noise (S/N)
spectra of small/medium stellar samples ($\sim$100-1000 stars) have been
used to unveil the chemo-dynamical properties of the different galactic
components (e.g., Gratton et al.  \citealp{gratton96,gratton00,gratton03},
Fuhrmann \citealp{fuhrmann98,fuhrmann08}, Adibekyan et al. 
\citealp{adibekyan}, Kordopatis et al., \citealp{kordopatis1}, 
Bensby \& Feltzing \citealp[and references therein]{bensby12}). 
Most of these studies have the disadvantage that they are
based on pre-selected samples defined according to strict
kinematical criteria and hence suffering from biases that are difficult to
quantify.

On the other hand, large spectroscopic surveys, such as the Geneva
Copenhagen Survey (GCS, Nordstr\"om et al.  \citealp{nordstrom04} --
$\sim$16000 stars), the Sloan Extension for Galactic
Understanding and Exploration\footnote{For recent results with SEGUE data see
Cheng et al., \citealp{cheng12} ($\sim$7000 stars), Schlesinger et al. 
\citealp{schlesinger11} ($\sim$40000 G- and 23000 K-dwarfs), Bovy et al. 
\citealp{bovy11} ($\sim$30000), Lee et al.  \citealp{lee11} ($\sim$17000
stars) and Liu \& van de Ven \citealp{liu12} ($\sim$27500 stars).} (SEGUE,
Yanny et al. \citealp{yanny}), and the RAdial Velocity Experiment (RAVE,
Steinmetz at al. \citealp{steinmetz}, Zwitter et al. \citealp{zwitter08},
Siebert et al. \citealp{siebert11})
take advantage of their large sample statistics (hence sampling a large
parameter space) to compensate for the much less precise measurements of the
abundances and stellar parameters, which are typical for medium/low
spectral resolution (or photometry, in the case of GCS).

Most of the surveys quoted above aim at estimating the [$\alpha$/Fe] ratios
of large samples of stars, which, to a first approximation, can be used as a
proxy for the stellar relative age (this seems to be valid even
when radial migration is present, see Sch\"onrich \&
Binney \citealp{schoenrich1}, Minchev, Chiappini \& Martig
\citealp{minchev12c}).
The uncertainty in the [$\alpha$/Fe] ratios
coming from the SEGUE abundance pipeline (Lee et al.,
\citealp{lee08a,lee08b}) are around 0.2~dex, for S/N above $\sim$~20.  For
the GCS much less precise estimates, based on Str\"omgren indices as proxy
for [$\alpha$/Fe] were obtained (Casagrande et al., \citealp{casagrande}).

Although the original goal of the RAVE survey is to obtain radial velocities
for up to a million stars, its spectral measurements turned out to be very
useful for chemo-dynamic analysis.  Recently, the first results of the RAVE
abundance pipeline were published (the RAVE Chemical Catalogue -- Boeche et
al., \citealp{boeche11}), showing that it is possible to measure up to
seven \emph{individual} elements from spectra with S/N above 40, and at
least three of them from spectra with S/N $\sim$~20.  The
RAVE-chemical pipeline
uncertainties for [$\alpha$/Fe] are similar to those of SEGUE -- around
0.2~dex.

A first study of the kinematic-abundance properties of the MW thin and
thick discs using RAVE data has been carried out by Karata{\c s} \& Klement
\cite{karatas12}, using results coming from the main RAVE pipeline (not from
the chemical pipeline). The authors used a sample of $\sim$4000
main-sequence stars (with log g $>$~3) from the second RAVE data release and
classified the Galactic disc populations according to their distribution on
the $V_{\rm rot}$-[M/H] plane, using the so-called X stellar population
parameter defined by Schuster et al.  \cite{schuster93}\footnote{This
parameter reflects the fact that thin disc stars tend to be more metal-rich
and faster rotating than stars in the thick disc, although it is clear that
both components strongly overlap both in metallicity and Galactic rotation
velocity.}.  Although the authors successfully reproduced the mean
kinematic properties of the thin- and thick-disc populations as compared to
previous work (e.g., Veltz et al., \citealp{veltz08}), their results should
be taken with caution.  Indeed, the metallicities and in particular the [$\alpha$/Fe]
estimates coming from the RAVE main pipeline are not free of considerable errors and systematical effects (see a
discussion in Siebert et al., \citealp{siebert11}), which might be a problem
for a classification method relying on metallicity, as the one adopted in
Karata{\c s} \& Klement \cite{karatas12}.

Recently, the RAVE project adopted a new pipeline
(Kordopatis et al. \citealp{kordopatis} and Kordopatis and the RAVE
collaborators, in preparation) to
estimate the stellar parameters values as effective temperature, gravity and
metallicity ($T_\mathrm{\rm eff}$, $\log g$, [M/H]), which is free of
most of the systematic errors cited before. Such values are required as input
to the RAVE chemical pipeline which estimates the chemical elemental
abundances.
To date, the RAVE-chemical catalogue constitutes the largest sample for
which individual chemical abundances are available.  Here, we exploit the
RAVE chemical catalogue  in the framework of abundance-kinematic
correlations of stars in the Galaxy.  
We aim at disentangling the different Galactic stellar components, as
well as at identifying
stars which were accreted or experienced heating and/or radial migration
along the evolution of our Galaxy. Possible processes that have been studied by means of N-body simulations include
stellar diffusion driven by transient spiral arms (Sellwood \& Binney
\citealp{sellwood02}, Ro{\v s}kar et al.  \citealp{roskar08}), by the interaction
between spiral arms and the Galactic bar (Minchev \& Famaey \citealp{mf10},
Minchev et al.  \citealp{minchev11a}, Brunetti et al.  \citealp{brunetti11})
or depositing stars into the Galactic discs by mergers (Abadi et al. \citealp{abadi}, 
Villalobos \& Helmi \citealp{villalobos}, Di Matteo et al. \citealp{dimatteo12}).  
Such a disc stirring can give rise to a number of phenomena in the
chemo-kinematical properties of a galaxy, such as flattening in radial
metallicity gradients (e.g., Sch\"onrich \&
Binney \citealp{schoenrich}, Minchev et al.  \citealp{minchev11a},
Pilkington et al.  \citealp{pilkington12}), extended stellar density
profiles (e.g., S{\'a}nchez-Bl{\'a}zquez et al.  \citealp{blazquez09},
Minchev et al.  \citealp{minchev12}), and can have
profound impact on the way we interpret the abundance-kinematic correlations
in our own Galaxy as recently discussed by Minchev, Chiappini \& Martig
\cite{minchev12c}.

From the observational side, the main difficulty has been the proper
disentanglement of the local thin and thick discs.  No selection criterion is
free from biases, be it kinematical or chemical (some drawbacks of a
separation based purely on the [$\alpha$/Fe] ratios in the framework of the
SEGUE sample are discussed in Brauer et al.  2012, in preparation).   
Recently Bovy et al \cite{bovy11} even argued that there is no clear 
dichotomy between the thin and the thick disc, a result in contrast to 
previous findings in similar studies (Veltz et al. \citealp{veltz08}).  It thus
seems that the best way to address this intricate problem is to study the
properties of a sample spanning the largest possible parameter space, both
in kinematics and in chemistry.

In  the present work we make such attempt by using the best quality data of
the RAVE abundance catalogue.  Our two primary goals are: a) to show that
the RAVE data give results consistent with those found from
higher-resolution, higher-S/N analysis for the chemo-dynamical properties of
nearby stars; and b) to properly investigate the abundance-kinematical
properties of the thin and thick discs with a sample 10 times larger,
and which covers a much larger volume (from $\sim$ 100~pc
to $\sim$ 3-4~kpc), than the one adopted in the pioneering work of
Gratton et al.  \cite{gratton03}, hereafter G03.

The layout of the paper is as follows: in Section 2 we discuss our sample selection. 
Section 3 describes how the
orbital parameters were computed.  In Section 4 we carry out the analysis of
the kinematic-abundance properties of the thin and thick discs divided
according to the kinematic selection criteria adopted in G03 and show
that the RAVE abundance pipeline gives trustworthy results, despite the medium
resolution of our spectra.
In Section 5 we propose a different approach to the problem of disentangling different
Galactic populations.  Finally, discussion and conclusions are
presented in Section 6.

\section{Data and Sample Selection}\label{stellarparam}
This work makes use of the full suite of the RAVE data products, namely radial 
velocities, stellar parameters, distance estimates and chemical abundances. 
Information on the radial velocities and proper motions come from 
Siebert et al. \cite{siebert11}, distances have been computed following the
recipe described in Burnett \& Binney \cite{burnett}.

Recently RAVE adopted a new pipeline (Kordopatis et al.,
\citealp{kordopatis}) for estimating the stellar parameters values. This
pipeline makes use of the codes MATISSE (Recio-Blanco et
al., \citealp{recioblanco}) and DEGAS (Bijaoui et al., \citealp{bijaoui})
optimized for the \ion{Ca}{ii} triplet region. 
These stellar parameters are
adopted for the present work and a more comprehensive description of the data
will be presented in the next RAVE data
release (Kordopatis and RAVE collaborators, in preparation).\\

The present chemical abundances internal data release employed for this 
study contains elemental abundances 
for 245649 MW stars (for a detailed description of how these abundances are obtained, see
Boeche et al. \citealp{boeche11}).
Roughly 10\% of the RAVE stars have been observed more than once. In order
to avoid using more than one estimation per star, for the re-observed stars
we here adopt only the values derived from the spectrum with highest
S/N.

For this first paper using the RAVE abundance catalogue, we select stars with the
highest quality spectra and abundances. We additionally require that such
stars cover a chemical-abundance range as large as possible (both in
metallicity and [$\alpha$/Fe] ratios). To satisfy the first condition, we
select spectra on which the code MATISSE converged to a single point of the
parameter space ($T_\mathrm{\rm eff}$,$\log g$, [M/H]), 
having high signal-to-noise ratio (at least S/N=60), which are well fit
by the reconstructed spectra of the chemical pipeline ($\chi^2<1000$) and
with no continuum defects\footnote{The $frac$ parameter
described in Boeche et al. \cite{boeche11} gives the fraction of pixels
that are non-defective and represents the goodness of the continuum fitting.} ($frac>0.99$). 
We also take care that the selected spectra are not peculiar (classified as
normal stars according to the three classification flags described in Matijevi{\v c}
et al. \citealp{matijevic}). To
satisfy the second condition, we select only cool 
giants and avoid dwarf stars.
Indeed, owing to their weaker lines, 
the abundance measurements in dwarfs are more uncertain, especially for
[X/H]~$<-1.0$~dex. Conversely, for cool giants, thanks to their intense absorption 
lines, the RAVE abundance pipeline is able to measure chemical elements down 
to [X/H]~$=-2.0$~dex. We exclude  giants with $\log g<0.5$, to avoid any
possible effects due to the boundaries of the learning grid used for the
automated parametrization (see Kordopatis and RAVE collaborators, in
preparation). 

In the present work we consider the chemical abundances of Fe and $\alpha$
elements (the latter obtained as average of [Mg/H] and [Si/H]).  These elements
have the most precise abundances of all the elements included in the catalogue
and, thus, are the most suitable ones for comparisons with the G03 work.  

We decided to work with two samples of different S/N: the main one contains only
spectra with S/N$>$75 (the ``SN75 sample") and the second one with S/N$>$60
(the ``SN60 sample"), used when a better statistic
is needed because it contains a larger number of stars\footnote{We verified that 
all the results of this work
are valid for both samples.}. Both samples satisfy the following criteria: i) giant
stars with gravity $0.5<\log g<3.5$; ii) effective temperature
$4000<T_\mathrm{\rm eff}$(K)$<5500$; iii) high quality data (spectra with
$\chi^2<1000$ and $frac>0.99$).  
With these criteria, the SN75 and the SN60 samples count 2167 and
9131 stars, respectively.

\begin{figure}[t]
\centering 
\resizebox{\hsize}{!}
{\includegraphics[width=9cm]{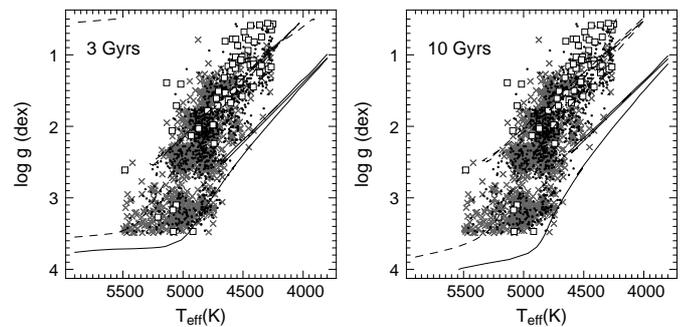}}
\caption{
$\log g$ versus $T_\mathrm{eff}$ values for the SN75 RAVE sample (see text) with 
overplotted isochrones of 3~Gyr (left panel) and 10~Gyr (right panel) with 
metallicities [M/H]=0.0~dex (solid lines) and [M/H]=--1.0 dex (dashed lines).
Black points, grey crosses and open squares represent the
thin disc, dissipative and accretion components as defined in
Sec.~\ref{kin_criteria}.
}
\label{isochrones} 
\end{figure}

\begin{figure*}[t]
\centering 
\resizebox{\hsize}{!}
{\includegraphics[clip=,width=16cm]{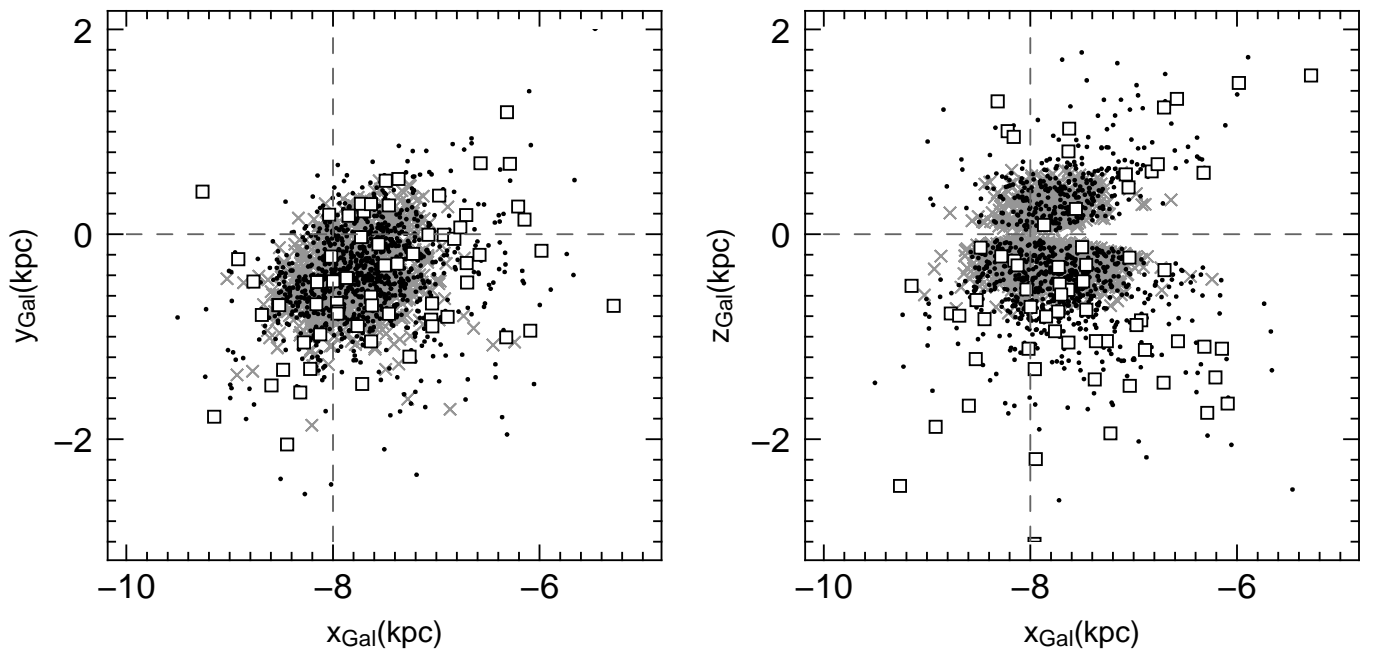}}
\caption{
Distribution of the SN75 RAVE sample in the $x_{\rm Gal}-y_{\rm Gal}$ and 
$x_{\rm Gal}-z_{\rm Gal}$ planes (left and right panels, respectively). The
dashed lines cross at the Sun's position. Symbols as in
Figure~\ref{isochrones}.
}
\label{xyzGal} 
\end{figure*}

These stellar parameter constraints
are optimal for selecting cold giant stars from the RAVE sample and avoiding
the hot giant stars of the horizontal branch. 
Figure~\ref{isochrones} shows where our sample
lies on the $T_\mathrm{eff}$-$\log g$ plane.  Also plotted are the
isochrones of Marigo et al.  \cite{marigo08} for metallicities 0.0 and
--1.0~dex and ages between 3 and 10~Gyr.  Figure~\ref{xyzGal} shows the
spatial distribution of our sample on the $x_{Gal}$-$y_{Gal}$ (Galactic
disc) and $x_{Gal}$-$z_{Gal}$ (vertical direction) planes, where the Sun's
location is indicated by the intersection of the dashed lines in each panel.

\section{Orbital parameters}

To be self-consistent with the new stellar parameters adopted
in this work (from the new RAVE pipeline - Kordopatis et al.  in prep.), we
recomputed the chemical abundances using the RAVE chemical pipeline
(Boeche et al.  \citealp{boeche11}) and new distances for our stars by using
the method described in Burnett \& Binney \cite{burnett} improved to take in
account interstellar extinction and reddening (Binney et al., in
preparation).
We next computed the 3D space velocities $u$, $v$ and $w$ along the
Galactic coordinates $x_{\rm Gal}$, $y_{\rm Gal}$ and $z_{\rm Gal}$.
We corrected the velocities for the local standard of rest velocity as
derived by Dehnen \& Binney \cite{dehnen98b}.
In order to obtain additional orbital parameters, we numerically integrated the orbits of 
stars by using the code NEMO (Teuben, \citealp{teuben}), given their stellar distances and velocities. 
For the Galactic potential we adopted the
model n.2 by Dehnen \& Binney \cite{dehnen98}, which assumes R$_0$=8.0~kpc, circular
velocity 
at the solar circle v$_c$(R$_0$)=217~km s$^{-1}$, and disc surface density 
$\Sigma$=52.1~M$_\odot$ pc$^{-2}$ (even when potentials
n.1, 3 or 4 by Dehnen \& Binney \cite{dehnen98} were employed, the results
did not change significantly). 
From the integrated Galactic orbits we extracted useful quantities, such as
apocentre, 
$R_a$, pericentre, $R_p$, eccentricity, $e$, Galactic rotation
velocity\footnote{Following the common use, the name ``Galactic rotation velocity"
refers here to the azimuthal velocity in a cilindrical coordinate system,
computed as $V_{\rm rot}=R\cdot \frac{d\phi}{dt}$.} $V_{\rm rot}$, and
the maximum vertical amplitude, $Z_{\rm max}$. 
$R_a$ and $R_p$ are the maximum and minimum distances from the Galactic 
centre a given star obtains during a revolution of $2\pi$ radians around the 
Galactic centre, measured on the Galactic plane. Similarly, $Z_{\rm max}$ is 
the maximum altitude reached by a star along its orbit. 
The eccentricity is defined as $e=(R_a-R_p)/(R_a+R_p)$.

\begin{figure}[t]
\centering 
\resizebox{\hsize}{!}
{\includegraphics[clip=,width=9cm]{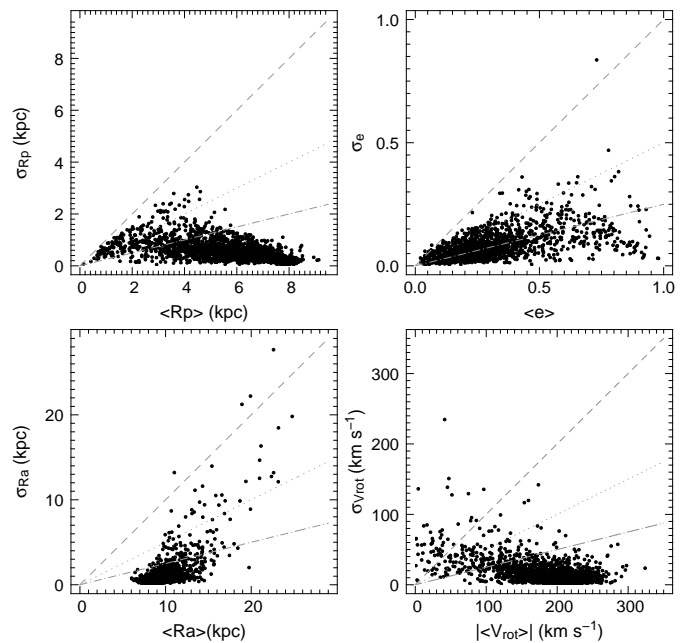}}
\caption{Estimated uncertainties versus their respective stellar orbital parameters value
for the SN75 sample.
Dashed, dotted and dash-dotted lines represent 100\%, 50\% and 25\% errors,
respectively.}
\label{mc_simul_stat} 
\end{figure}

In order to estimate errors in the orbital parameters we followed the
G03's method, i.e., we performed a Monte Carlo simulation: for each star we
computed 100 times the orbital parameters by changing every time the
Galactic coordinates $x_{\rm Gal}$, $y_{\rm Gal}$, $z_{\rm Gal}$ and the 
velocities $u$, $v$, $w$, according to their estimated errors. From 
these 100 orbits we computed the standard deviation for each orbital
parameter. Despite the large distance for some of our stars (up to 3~kpc, 
see Figure~\ref{xyzGal} and Figure~\ref{figToomre} top-right panel) for which
the proper motions (and therefore the tangential velocities) are very uncertain,
the orbital parameters show reasonably small variations when errors are taken
in account.
As showed in Figure~\ref{mc_simul_stat}, the errors in eccentricity
are smaller than 0.2 for most stars, while for $R_p$ they are
smaller than 1~kpc. Similar behaviour is seen in the other parameters.
With such errors we are confident that we can use these orbital parameters to
discriminate between thin- and thick-disc stars, with only moderate
contamination.\\

During the preparation of this manuscript we used four different versions of
distance estimations (two preliminary and to definitive, following the
methods of Zwitter et al. \citealp{zwitter10} and
Binney et al. in preparation) and two different chemical catalog
versions (DR3 by Siebert et al. \citealp{siebert11}, and DR4 by Kordopatis et
al. in preparation) for our analysis. The results of the present work
(see next sections) hold for all the data set we employed, confirming 
the robustness of the selected sample and the results obtained.

\section{The kinematical criteria of G03 applied to the RAVE
data}\label{kin_criteria}

In G03 the authors used 150
field subdwarfs and early subgiants with accurate parallaxes and kinematics 
and divided them into three subsamples according to pure kinematic criteria.
With the purpose of validating the RAVE abundance pipeline, as well as our orbital parameters, we follow 
here the analysis done in G03 
and divide the RAVE samples into three subsamples according to the following criteria:

\begin{enumerate}
\item The {\em thin disc component} including stars whose orbits have low eccentricity and low maximum
altitude from the Galactic plane, i.e. $e<0.25$ and $Z_{\rm max}<$ 0.8~kpc. 
It will be represented as grey crosses in the figures that follow.

\item The {\em dissipative collapse component} consisting of stars with $V_{\rm
rot}>40$~km~s$^{-1}$. Stars belonging to the thin disc component just defined
are excluded.
This sample includes part of the thick disc, as well as of the halo. Through
all the figures this sub-sample will be represented with black dots;

\item The {\em accretion component}
of non-rotating or counter-rotating stars. These stars satisfy the
constraint $V_{\rm rot}\le40$~km~s$^{-1}$.  This population is very
likely composed both from halo stars and accreted debris
(i.e. stars that do not share the rotation of the disc components).  
It is represented in the figures as open squares.

\end{enumerate}

The criteria mentioned above deviate somewhat from the criteria used in G03, 
who defined the thin disc component via $\sqrt{Z_{\rm max}^2+4e^2} <0.35$ and 
had an additional constraint on the apogalactic radius $R_a <15$~kpc 
for the dissipative component. The thin disc stars selected 
by this constraint cannot have $e>0.175$ if they lie on the Galactic plane and they 
must have circular orbits if their $Z_{\rm max}=0.35$~kpc. Indeed, while 
the above criteria led to differences in the chemical composition that likely
reflect real differences in the stellar populations (as will be shown in
the next sections), the detailed specification is to some extent arbitrary, as already pointed out by 
 G03. Our modification allows us to generalize 
the selection criteria in the $e-Z_{\rm max}$ plane, as we will show 
in the next section. Furthermore, our sample covers a considerably larger
volume than the Hipparcos sphere probed by G03, resulting in a
relatively low number of stars with near-circular and coplanar 
orbits when the G03 original selection criteria is employed.

Indeed, the G03 sample is composed of dwarf stars, covering
a small spatial volume ($<100$~pc from the Sun) in order to include 
only stars with accurate parallaxes. Moreover, their sample was drawn from the Hipparcos catalogue for which
metal-poor stars were preferentially selected (with [Fe/H] $<$--0.8~dex).  Hence, 
this sample includes a rather large number of high-proper-motion stars, 
resulting in a strong kinematical bias favouring objects on highly eccentric 
orbits, with low Galactic rotation velocities, and with either large apogalactic or small
perigalactic distances.

The RAVE sample considered here, on the other hand, is composed by giant
stars, thus, probing a large volume of space (up to 3 kpc from the
Sun).  Most importantly, the RAVE sample is rich in high
metallicity ($>-1.0$~dex) stars.  Moreover, while the G03 sample is
composed of different high-resolution spectroscopic subsamples, partly
obtained by the authors and partly found in the literature, the RAVE sample
adopted here is very homogeneous.  Finally, our samples SN75 and SN60
have respectively $\sim$14 and $\sim$60 times the size of the G03 sample.

\begin{figure}[t]
\centering 
{\includegraphics[bb=292 502 468 670,clip=,width=9cm]{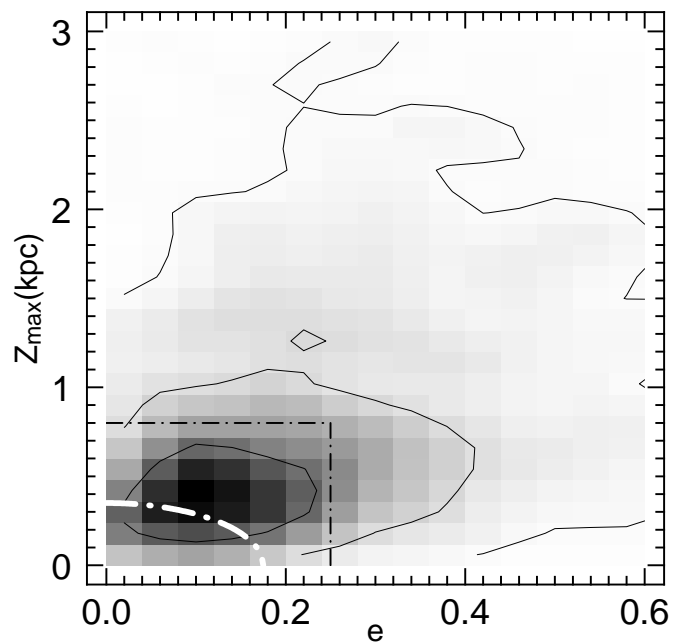}}
\caption{
Density distribution on the $e-Z_{\rm max}$ plane for our sample
SN75 sample, with 2152 stars. The dash-dotted
white curve shows the G03 thin-disc selection criteria, whereas the dashed-dotted 
black line frames the area defined by our modification to the $e$ and $Z_{\rm max}$
constraints. The contours contain 34\%, 68\% and 95\% of the sample.  
}
\label{Zmax_e_select} 
\end{figure}

In Figure~\ref{Zmax_e_select} we show number density contours 
of the maximum height achieved above the Galactic 
plane, $Z_{\rm max}$ versus the orbital
eccentricities, $e$ for the sample SN75.  The criterion $\sqrt{Z_{\rm max}^2+4e^2}<0.35$ 
adopted by G03 to define the thin-disc sample is represented by the quarter of an 
ellipse on the $e-Z_{\rm max}$ plane (white dash-dotted curve), the thin disc defined 
by our criterion is shown as black dash-dotted line.

\begin{figure*}[t]
\centering 
{\includegraphics[clip=,width=12cm]{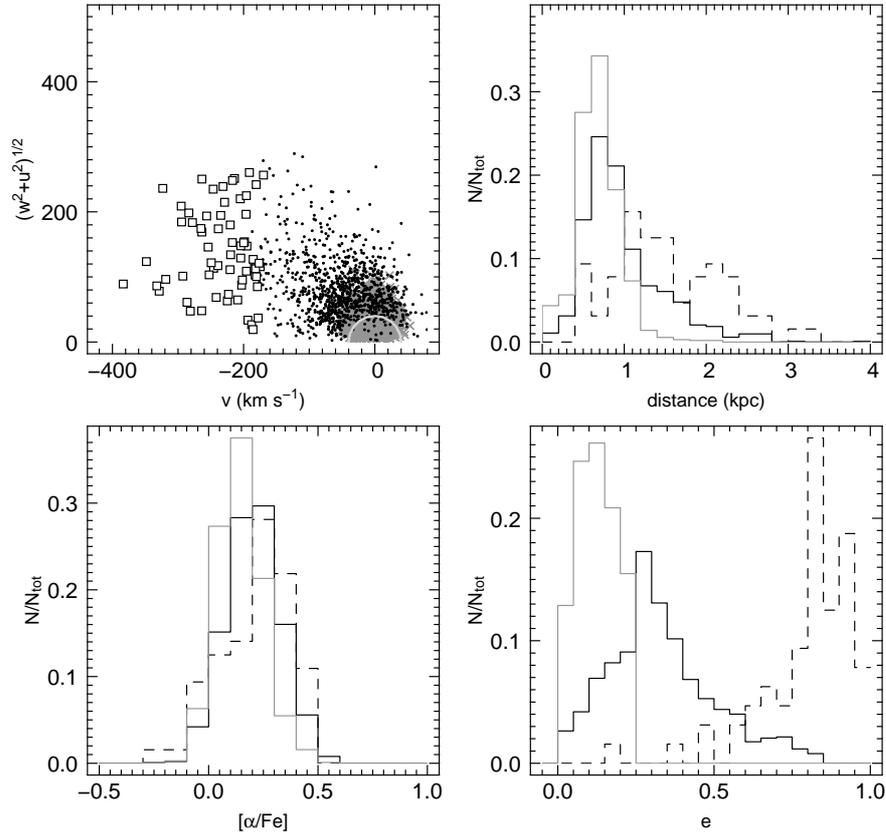}}
\caption{Toomre diagram (upper left) for each of our samples: thin disc
stars (grey points), dissipative component (mostly thick disc stars - black
dots) and accreted component (squares). The light grey semi-circle indicates 
the constant peculiar velocity $(u^2+v^2+w^2)^{1/2}=40$
km s$^{-1}$.
Also shown are the distance (upper
right), [$\alpha$/Fe] (bottom left) and eccentricity (bottom
right) distributions of the thin disc (grey line) dissipative
(solid black line) and accreted (black dashed line) components.} \label{figToomre}
\end{figure*}

\begin{figure}[t]
\centering
\resizebox{\hsize}{!}
{\includegraphics[clip=,width=9cm]{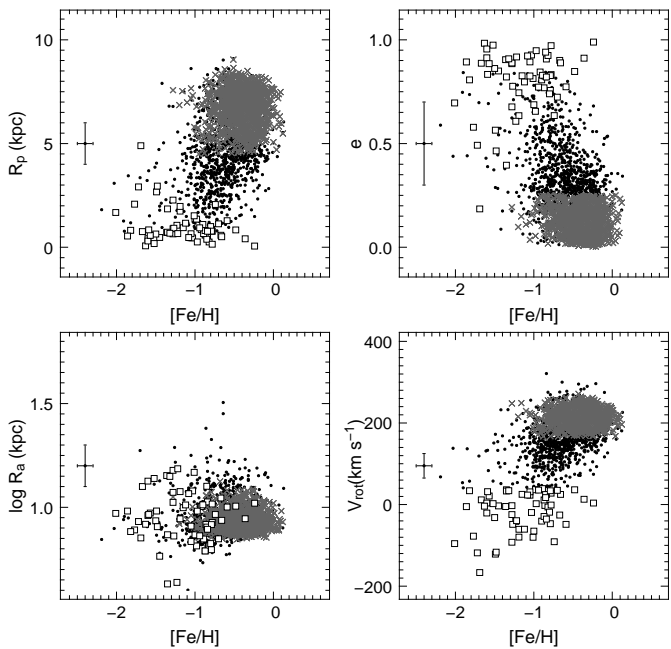}}
\caption{Perigalacticon, $R_p$, eccentricity, $e$, apogalacticon, $R_a$, and
Galactic rotation velocity, $V_{\rm rot}$ as a function of [Fe/H] for the three
subsamples. Symbols are as in Figure~\ref{figToomre}. This figure
corresponds to Figure~4 of G03.}
\label{fig4m} 
\end{figure}

\begin{figure}[t]
\centering
\resizebox{\hsize}{!}
{\includegraphics[clip=,width=9cm]{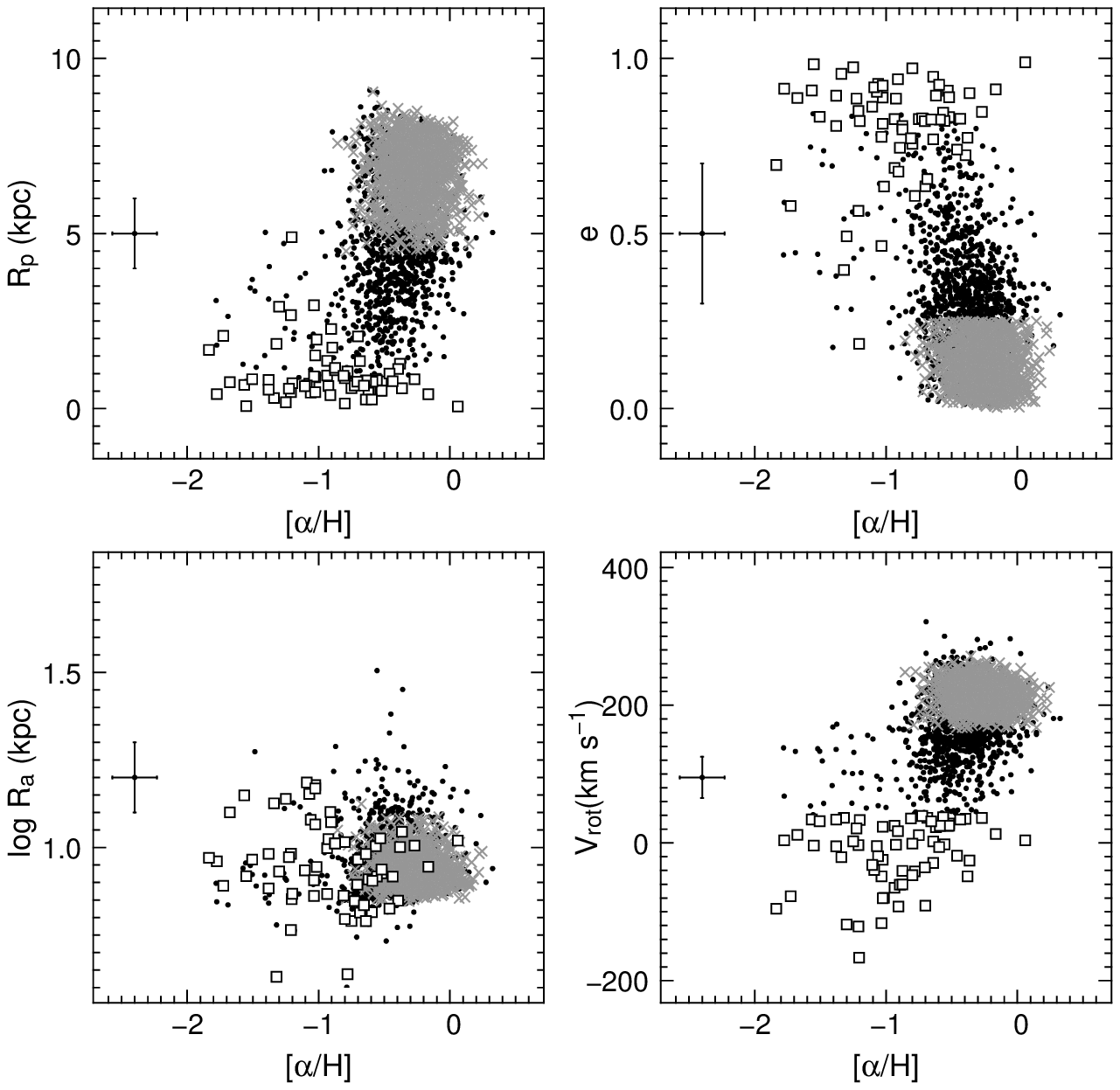}}
\caption{As in Figure~\ref{fig4m}.
This figure corresponds to Figure~5 of G03.}
\label{fig5m} 
\end{figure}

\begin{figure}[t]
\centering 
\resizebox{\hsize}{!}
{\includegraphics[clip=,width=9cm]{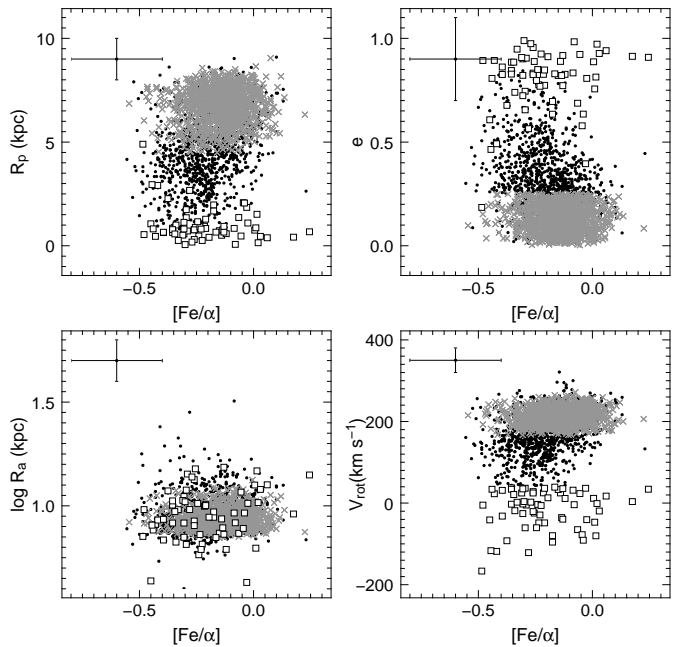}}
\caption{As in Figure~\ref{fig4m}. 
This figure corresponds to Figure~6 of G03.}
\label{fig6m} 
\end{figure}

With the above prescriptions, the thin-disc component consists of
1079 stars, the dissipative component of 
1024 stars, and the accretion component
of 64 stars. The
results of this division are shown in
Figures~\ref{isochrones}, \ref{xyzGal}, \ref{figToomre},~\ref{fig4m},~\ref{fig5m} and ~\ref{fig6m}. 
The thin disc, dissipative and accretion components show different chemical
signatures, similar to the ones seen in G03, with the difference that our sample 
extends up to solar abundances.  

Figure~\ref{figToomre} shows the main properties of the different
samples.
While the accretion component appear scattered over a wide area of the Toomre
diagram (and it would extend to higher $v$, since such non-rotating
population must hold rotating and counter-rotating stars in equal amount)
the thin disc and dissipative components clump and overlap on a smaller
area.
With respect to the distances, the outcome of our sample selection is that the
thin disc stars cover a smaller volume than the dissipative sample, whereas
the accreted stars prefer larger distances (between 0 and 3 kpc) from
the Sun.  The eccentricity
distribution of each sample is also shown.  It is interesting to notice that
the eccentricity and distance distributions found here are qualitatively 
similar to those found with the SEGUE G-dwarf sample used in Lee et al.  \cite{lee11}, even though
the latter authors adopted a pure chemical criterion to divide their sample
into thin- and thick-disc stars and even though our distance and eccentricity 
distribution is likely to be affected by our selection criteria.

Indeed, the RAVE giant sample
covers essentially the same volume as the SEGUE dwarf sample (Steinmetz
\citealp{steinmetz12}).
The similarities between our thin and dissipative samples and the thin and
thick disc ones obtained by Lee et al.  \cite{lee11}, is by itself
reassuring.  Indeed, two completely different surveys, using different tracers
(giants vs.  dwarfs), with chemistry, distances and orbital parameters
computed by different pipelines, analyzed in completely different ways
(kinematically selected vs. chemically selected), still give very similar answers
for the mix of Galactic stellar populations within $\sim$2--3kpc from the Sun.

In Figure~\ref{figToomre} we also show the [$\alpha$/Fe] distributions
for each of our samples.  An offset of $\sim$0.1 dex is seen between
the thin and dissipative samples, with however a considerable overlap (as
expected even according to pure chemical evolution models, where the difference in
[$\alpha$/Fe] between the thick and thin discs is a function of metallicity
- see Chiappini \citealp{chiappini09}).  Note that the absolute values of
the mean [$\alpha$/Fe] ratios of the thin disc and thick disc (dissipative)
distributions are in good agreement with the mean values reported by
high-resolution studies (Bensby and Feltzing \citealp{bensby12} and
references therein).  These values are, however, lower than the mean values
obtained by Lee et al.  \cite{lee11} with the SEGUE sample, namely:
[$\alpha$/Fe] $\sim$0.1 for the thin disc, and $\sim$0.35 for the thick disc
(a detailed comparison between RAVE and SEGUE results is
beyond the scope of the present paper, and will be presented by Brauer et
al.  in preparation).  

\begin{figure*}
\begin{minipage}[t]{9cm}
{\includegraphics[clip=,width=9cm]{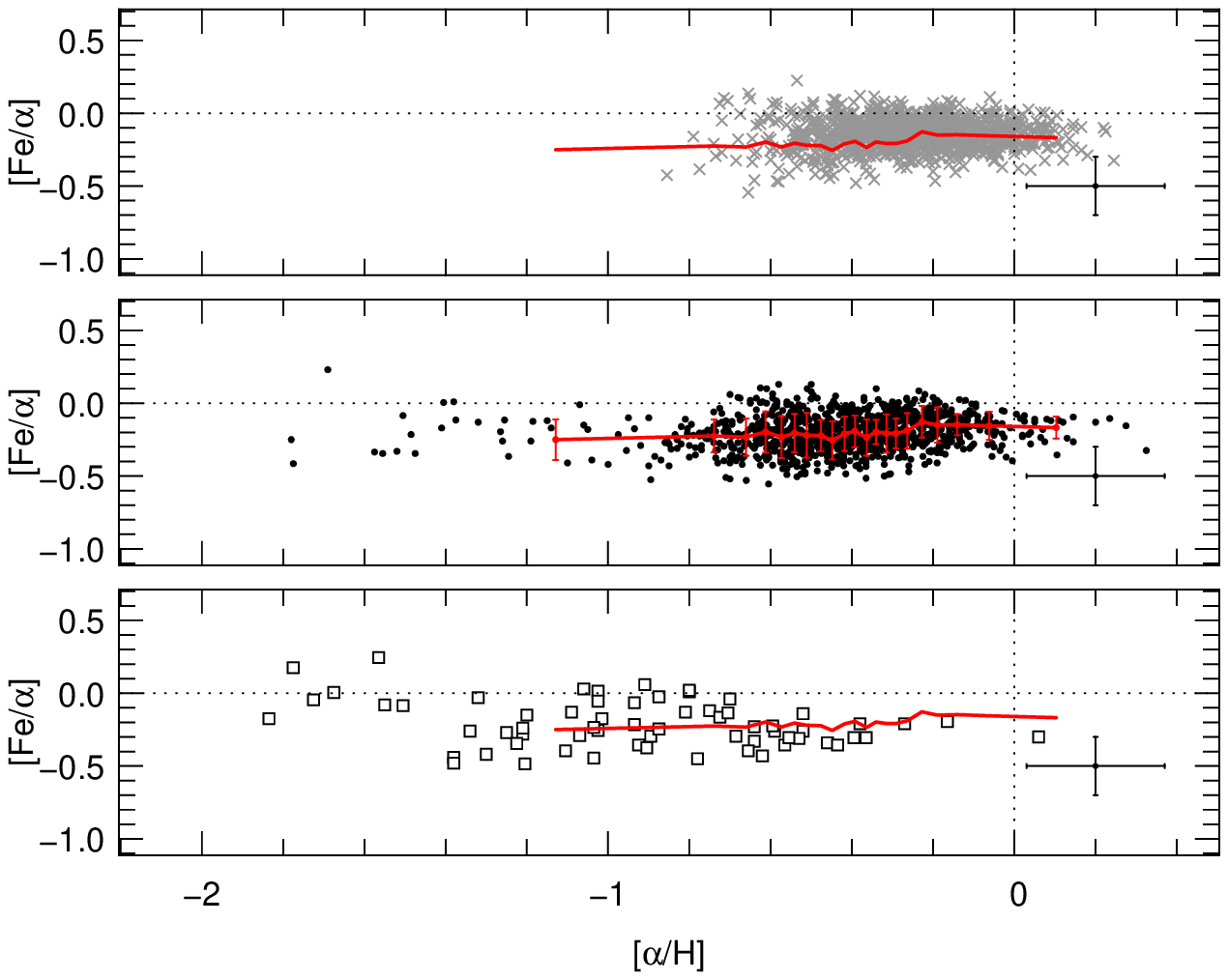}}
\caption{Abundance ratio [Fe/$\alpha$] versus the abundance  [$\alpha$/H] 
for the thin disc component
(grey crosses, top panel), the dissipative component (black points, middle panel), 
and the accretion component (open squares, bottom panel), selected by using our
modified criteria (see text). The red line
represents the average [Fe/$\alpha$] of the dissipative component obtained by
averaging bins of 50 points each, and the error bars represent their standard
deviation. The red line is reproduced in each panel
as a fiducial line. This figure corresponds to Figure~2 of G03.}
\label{fig2m} 
\end{minipage}
\hfill
\begin{minipage}[t]{9cm}
{\includegraphics[clip=,width=9cm]{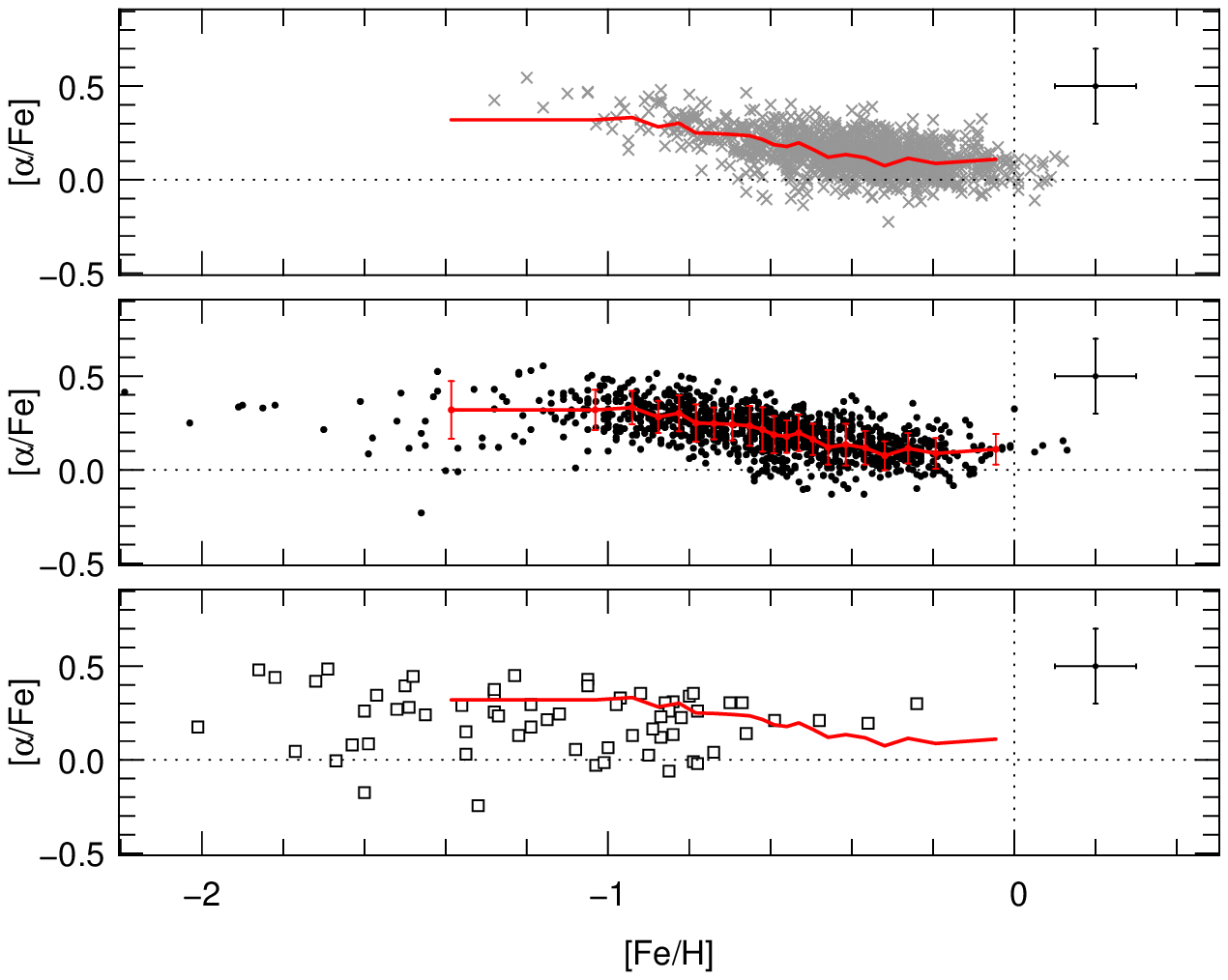}}
\caption{As in Figure~\ref{fig2m} but for the abundance ratio [$\alpha$/Fe] now as
function of [Fe/H]. This figure correspond to Figure~3 of G03.}
\label{fig3m} 
\end{minipage}
\end{figure*}

In Figures~\ref{fig4m}, \ref{fig5m}, and \ref{fig6m} we successfully
reproduced G03's Figures~4, 5 and 6, respectively, with some additional
features.  The RAVE data support some of the G03 results, such as the
existence of correlations between [Fe/H] and $R_p$, $e$, $V_{\rm rot}$, and
the apparent absence of a correlation with $R_a$
(Figure~\ref{fig4m}).  
The dissipative
component (black dots) has a wide range of eccentricity, and moderately high
abundances.  Also seen in Figures~\ref{fig4m} and \ref{fig5m} is an apparent drop in the
black point density around [Fe/H]$\sim -$1.0~dex and
[$\alpha$/H]$\sim -$0.7~dex (see also Figure~\ref{fig2m} and
Figure~\ref{fig3m} top and middle panels).
The standard deviation of the black points is
$\sigma_{[\alpha/{\rm H}]}$=0.2~dex (after the rejection of the accretion
component stars polluting the dissipative component at $[\alpha/{\rm H}]<$-1.0),
which is very close to the error in abundance expected for [$\alpha$/H] (0.17dex). 
This support the weak correlation shown with $V_{\rm rot}$ 
and --0.7~dex may represent the lower limit of the thick
disk $\alpha$-abundance.\\

In Figures~\ref{fig2m} and \ref{fig3m} we show the trend of [Fe/$\alpha$]
with [$\alpha$/H], and [$\alpha$/Fe] with [Fe/H] (similar to Figures 2 and 3 of G03, respectively). 

Both the
thin and dissipative components show a decreasing [$\alpha$/Fe] abundance ratio
with increasing metallicity.  However, the thin-disc component shows
systematically larger [Fe/$\alpha$] ratios than the dissipative one
for [$\alpha$/H]$<-0.2$~dex.  This can be
seen by the solid red line, which indicates the average of the black points
in bins with variable size in [Fe/H] so that every bin contains 50 points. 
By plotting this fiducial in each panel, the difference between the
thin-disc and dissipative component is visible in Figure~\ref{fig2m}. 
For [$\alpha$/H]$>-0.2$~dex the thin and dissipative components
seem to be chemically indistinguishable.
Note that for the accreted component the abundance ratio remains essentially
flat, and more importantly, shows systematically lower [$\alpha$/Fe] than the
dissipative component (see bottom panel of Figure~\ref{fig3m}). This result is not
only in agreement with the original findings of G03, but is also similar to
the recent high-resolution results by Nissen \& Schuster \cite{nissen}. 

It is clear that although the G03 criteria lead to clear
sub-populations, overlaps between the samples still exist. The $\alpha$-enhanced
stars at [Fe/H]~$<-0.7$~dex of the thin-disc component (Figure~\ref{fig3m},
top panel) probably belong to the dissipative component.
On the other hand, the stars at
[Fe/H]$<-1.0$~dex of the dissipative component are very likely accretion
component stars. In fact, because the accretion component stars are non-rotating (they
can be identified as halo stars by looking at the locus
they occupy in the Toomre diagram, top-left panel of \ref{figToomre})
their average $V_{\rm rot}$ must be zero.
By mirroring the square points with respect to $V_{\rm rot}=0$ ($v=-217$ km
s$^{-1}$), we can infer that many black points (dissipative component) could
also be considered accretion component stars.

Despite of the small overlaps described above, the three groups of stars
assigned to the different components via kinematical criteria, show
different chemistry:

\begin{itemize}
\item {\em thin disc component}: the distributions in $R_a$, $R_p$, $e$ 
and $V_{\rm rot}$ span a small range of values because they are limited by 
the criteria $e<0.25$ (nearly circular orbits) and $Z_{\rm max}<0.8$~kpc  
(close to the Galactic plane), i.e., stars are contained in a local volume 
with $R=8\pm1$~kpc, see Figure~\ref{xyzGal}. 
This component has an [$\alpha$/Fe] peak slightly above zero
and tend to be more shifted
to higher metallicities (figures \ref{fig4m} and \ref{fig5m}).

\item {\em dissipative component}: for this subsample $V_{\rm rot}$ is 
downward limited by the constraint $V_{\rm rot}>$~40~km s$^{-1}$ and stars which 
happen to have circular orbits ($e<$~0.25) are missing from this component in 
favour to the thin disc component. This creates a small bias against the 
dissipative component which diminishes the number of stars at low
eccentricities,
but does not completely remove them, since the orbital parameters can still span a wide 
range of values for $R_p$ and $e$ as well. On the other hand, stars at 
[Fe/H]$<-1.0$~dex are likely to be members of the accretion component.
The dissipative component has more stars at high [$\alpha$/Fe] 
with respect to the thin-disc component. It also shows a correlation between
the abundances [Fe/H], [$\alpha$/H] and the orbital parameters, as found also by G03.
\item {\em accretion component}: 
the kinematical criteria adopted for this sample imply that such
objects have mainly high eccentricity (Figure~\ref{figToomre}). 
This component covers a wide
range in [Fe/H], including objects with metallicities clearly larger than
the upper limit halo stars metallicity of $\sim -1$.

\end{itemize}

The successful reproduction of the main results of G03 validates the
kinematical and chemical data of the RAVE survey (despite the fact that they
were obtained from medium-resolution spectroscopy)
and allows us to push further our analysis.  

However, any selection criteria aiming at disentangling the thin and thick discs will suffer 
from the fact that these two components overlap in almost all parameters.
Pure kinematical selection criteria of thin and thick disc return
samples which partially overlap in chemical abundances (Bensby et al. \citealp{bensby03},
\citealp{bensby05}, Reddy et al. \citealp{reddy06}, among others), whereas
pure chemical selection criteria return samples which partially overlap in
kinematics (Navarro et al. \citealp{navarro11}, Lee et al.
\citealp{lee11}).
Even if a clear kinematical separation between these two components did
exist in the past, it could have been heavily blurred by a number of agents. 
In the next Section we try an alternative approach with the aim of avoiding
strict selection criteria, which hopefully can be more successful in
providing more robust constraints to chemo-dynamical models.

\section{The $e-Z_{\rm max}$ plane}\label{e-Zmax_sec}

As discussed in Section 1, there exist two 
general processes, resulting from the disc secular evolution, 
which could partially destroy the kinematical borders between 
the thin and thick discs: (1) kinematical heating (increase of stellar velocity 
dispersion) with time and (2) radial migration (change in the angular momentum 
of stars and, thus, in their guiding radii). Another possibility is (3) the
deposition of material by accretion, making the situation even more complex
since mergers also can give rise to (1) and (2) above. Processes (1), (2)
and (3) will lead to different signatures:

(1) Heating by transient (Carlberg \& Sellwood, \citealp{carlberg85}) and
multiple (Minchev \& Quillen, \citealp{minchev06}) spiral density waves, the
Galactic bar (Minchev \& Famaey \citealp{mf10}), giant molecular clouds
(Jenkins \& Binney \citealp{jenkins90}), and minor merger events (Quinn et
al.  \citealp{quinn93}) would most probably increase\footnote{Note, however,
that this effect still needs to be better quantified for a MW-like galaxy.}
the velocity dispersions of all three components ($U$, $V$, and $W$). An
increase in the vertical velocity dispersion would then result in
overlapping an initially cooler stellar population (such as the thin disc),
for which the heating is more effective due to the cold orbits, with an
initially hotter stellar sample (such as an old thick disc), which would not
be affected much by the heating agents.  
Note that the above would be true irrespective of whether the thick disc
were born hot (Forbes et al. \citealp{forbes}) or were preheated by mergers at high
redshift (e.g., Villalobos and Helmi \citealp{villalobos}) by virtue of the thin disc being
younger than the thick disc and the expected decrease of merger activity
with redshift.

(2) In contrast, as a star migrates (gains or loses angular momentum),
information about its birth radius is lost and its kinematics can be
virtually indistinguishable from those of stars born at the new radius
(Sellwood \& Binney \citealp{sellwood02}, Minchev et al.
\citealp{minchev11b},\citealp{minchev12b}).  In
recent years radial migration (or mixing) has been recognized as an
important process affecting galactic discs.  Several radial migration
mechanisms have been described in the literature: the effect of the
corotation resonance of transient spiral density waves (Sellwood \& Binney
\citealp{sellwood02}, Ro{\v s}kar et al.  \citealp{roskar08}, Sch\"onrich \&
Binney \citealp{schoenrich}), the effect of
the non-linear coupling between multiple spiral waves (Minchev \& Quillen,
\citealp{minchev06}) or bar and spirals (Minchev et al. 
\citealp{mf10,minchev11a}, Brunetti et al.  \citealp{brunetti11}), and the
effect of minor mergers (Quillen et al.  \citealp{quillen09}, Bird et al. 
\citealp{bird11}).

(3) Deposition of material by accretion has been shown to commonly occur in
cosmological simulations (e.g., Abadi et al. \citealp{abadi}). In this scenario tidal
debris of satellites with orbits coinciding with the plane of the host disc
can populate a disc component and, thus, blur the borders between the
preexisting thin and thick disc.

Such diffusion mechanisms make ineffective any kinematic criteria aiming at separating
the thin from the thick disc, even in the unrealistic case of no errors in
the kinematical data available.  The chemical properties acquired at
birth, on the other hand, must be preserved.  However, here the difficulty
is that due to the shallow abundance gradient in the galactic disc, as well
as the small range in the [$\alpha$/Fe] variation among different galactic
populations (at most 0.5 dex), the chemical differences are subtle (unless
one uses abundance ratios involving other chemical elements which present
larger variations).  In addition, even if the differences in the star
formation histories of the thick and thin discs would have led to chemical
differences (see Chiappini \citealp{chiappini09}), large overlaps would
still exist. This would remain true if the disk was composed of several
mono-abundance sub-components, as suggested
by Bovy et al. \cite{bovy11}.
For these reasons, we here abandon the ``selection criteria"
approach and try to have a different view of the problem.

Our previous analysis considered the distributions of the orbital parameters 
(such as $Z_{\rm max}$, $R_p$, $e$ and $V_{\rm rot}$) separately. However, 
these distributions are slices in a bigger chemo-kinematical, multidimensional 
space, in which the stars lie. It is therefore possible that some information could 
be missed when only the shapes of these distributions are considered, resulting
in a mixup in our separation of different Galactic components.
For instance, consider two stars 
with $V_{\rm rot} \sim$220 km s$^{-1}$: one can have a
circular orbit on the Galactic plane at the Sun radius $R_0$,
and the other can have $R_p<R_0$, an eccentric orbit
and vertical velocity $v_z$ large enough to reach $Z_{\rm max}>2$ kpc.
It could be hardly thought that such stars belong to the same population.
To avoid such a trap, we further analyse our sample by grouping
the stars by similar orbits using the $e-Z_{\rm max}$ plane. The eccentricity 
gives the shape of the orbits, whereas $Z_{\rm max}$ tells us about the oscillation 
of the star perpendicularly to the Galactic plane.

\begin{figure}[t]
{\includegraphics[bb=99 496 280 670,clip=,width=9cm]{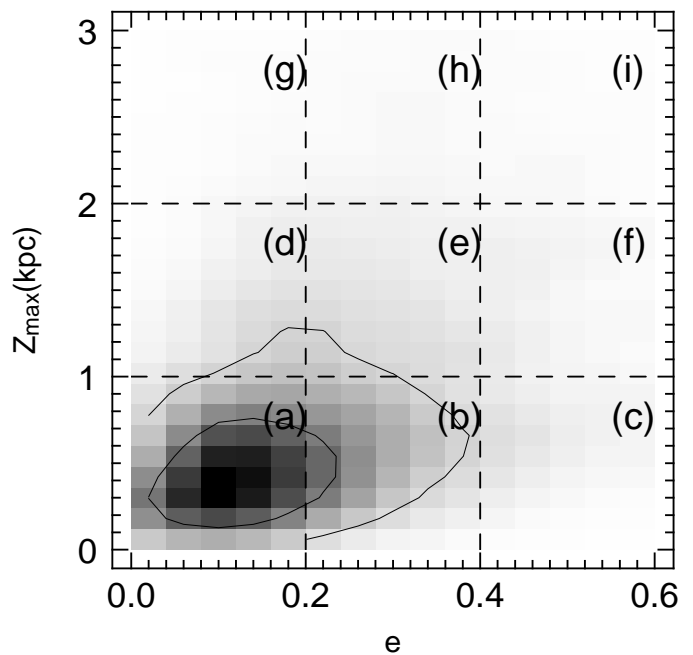}}
\caption{
The $e-Z_{\rm max}$ plane divided in nine panels labelled from (a) to (i).
Here we use the SN60 sample, which counts 9131 stars. 
}
\label{Zmax_e_sections} 
\end{figure}

In Figure~\ref{Zmax_e_sections} we divide the $e-Z_{\rm max}$ plane in 
nine groups of stars, as indicated by the dashed lines, and we label them from (a) to (i).
We have neglected stars with $e>0.6$ because here we focus on the Galactic disc. 
In this way we have sorted
the stars into ``classes" of orbits: moving rightwards the eccentricity
grows, moving upwards $Z_{\rm max}$ grows (hence the
vertical $v_z$ velocity increases). 

By dividing the $e-Z_{\rm max}$ plane into nine regions, we have obtained stellar 
subsamples with narrow ranges in orbital parameters. For each group we now plot
the [$\alpha$/Fe] versus [Fe/H] relation and 
the distributions of [Fe/H],  [$\alpha$/Fe], $R_p$, $R_m$, and $V_{\rm rot}$,  
in Figures~\ref{Zmax_e_aFe} and \ref{Zmax_e_Vphi}, respectively.
Panels (a), (c) and (g) of the aforementioned figures give  particularly interesting 
insights, on which we now focus.

\subsection{Identification of thin/thick/diffused stars} 
In the
following analysis we use the SN60 sample, in order to have a better
statistic\footnote{We verified that the results found in this work are valid
and consistent for both SN75 and SN60 samples.}. 
We first study the
distributions of subsample (a), as defined in Figure~\ref{Zmax_e_sections}. 
The stars in this group (with eccentricities $<$~0.2 and low
vertical velocities) show the expected properties of a sample dominated by
local thin-disc stars, namely: a Fe-distribution peak at $\sim
-$0.25~dex, a $V_{\rm rot}$
peak at $\sim$220~km~s$^{-1}$, and a mean galactocentric distance $R_m$ of
$\sim$7.5~kpc. The apparent low abundance of the Fe-distribution 
(which peaks at $\sim-$0.25~dex instead of 
$\sim-0.05$~dex of the local thin disk found by Casagrande et al.
\citealp{casagrande}) is due to the spatial distribution of the RAVE
sample, which lacks of nearby stars and favours stars lying at $z_{\rm
Gal}>300$~pc, where the metallicity
distribution function is shifted to lower values (also found by
Schlesinger et al. \citealp{schlesinger11} and consistent with the
predictions of the chemodynamical model
of Minchev et al. \citealp{minchev12c}). Nonetheless, the Fe-distribution in panel (c) is
more metal rich in comparison with the subsamples of the other panels.

\begin{figure*}
\begin{minipage}[t]{13cm}
\centering 
\resizebox{\hsize}{!}
{\includegraphics[clip=,width=14cm]{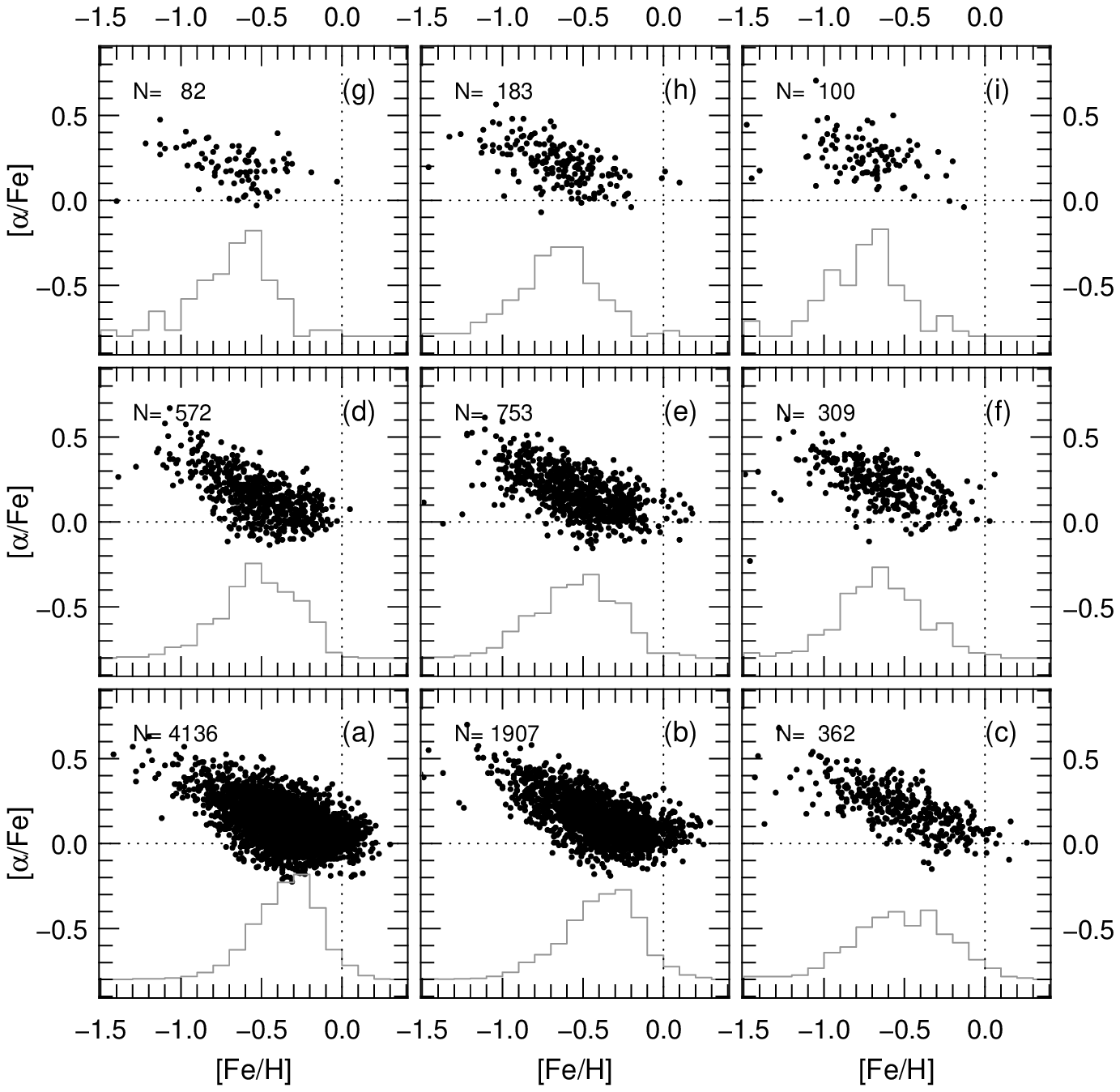}}
\resizebox{\hsize}{!}
{\includegraphics[clip=,width=14cm]{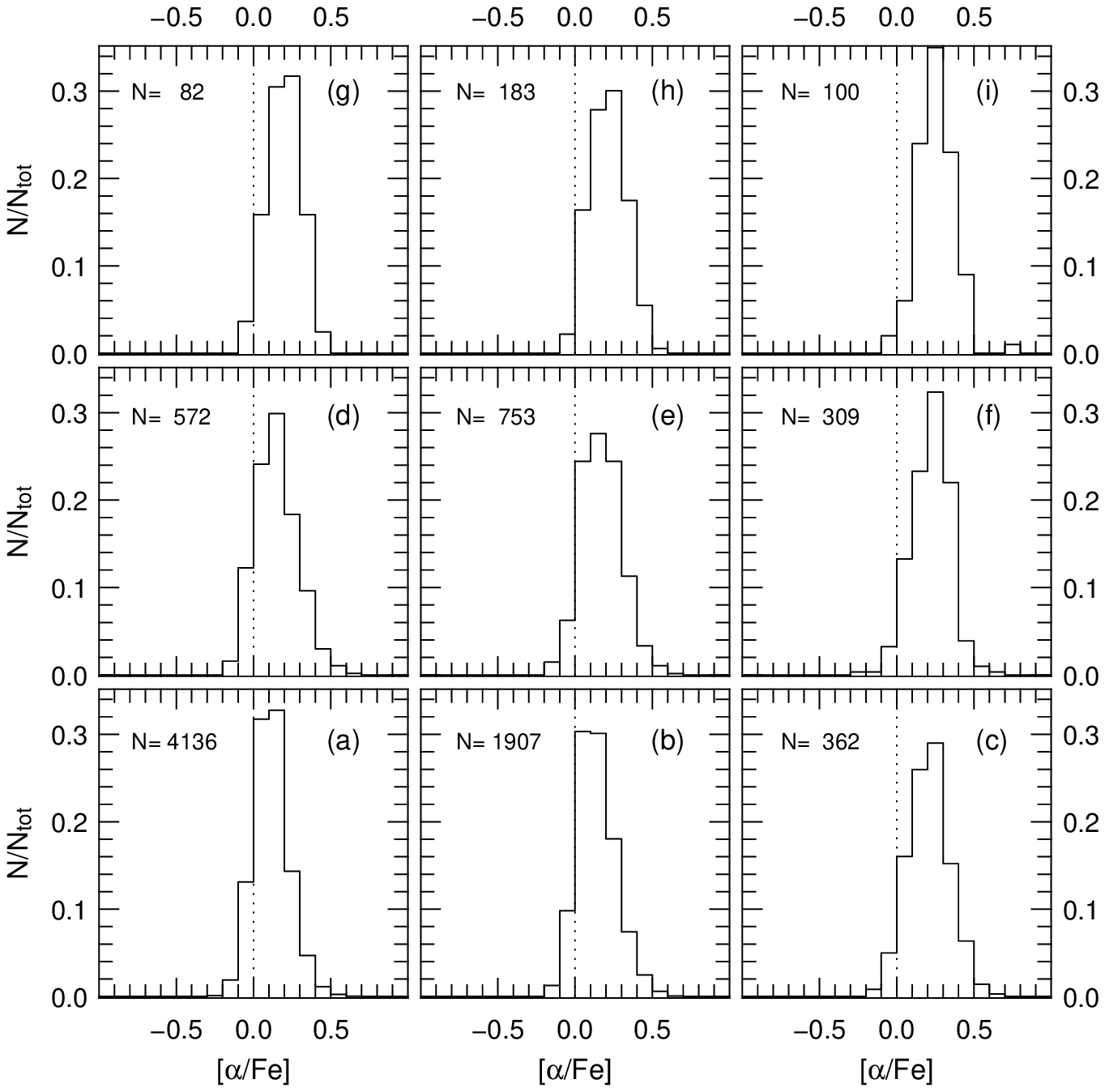}}
\caption{Upper panel: Relative abundance [$\alpha$/Fe] versus [Fe/H] for the stellar samples defined by 
panels (a) through (i) in Figure~\ref{Zmax_e_sections}. The histograms represent
the Fe distributions with relative scales. Lower panel: Distributions of abundance [$\alpha$/Fe] for the stellar samples defined by 
panels (a) through (i) in Figure~\ref{Zmax_e_sections}. The distributions are
normalized over the total number of points contained in each panel 
(N$_{\rm tot}$).}
\label{Zmax_e_aFe} 
\end{minipage}
\end{figure*}

\begin{figure*}
\begin{minipage}[t]{13cm}
\centering 
\resizebox{\hsize}{!}
{\includegraphics[clip=,width=14cm]{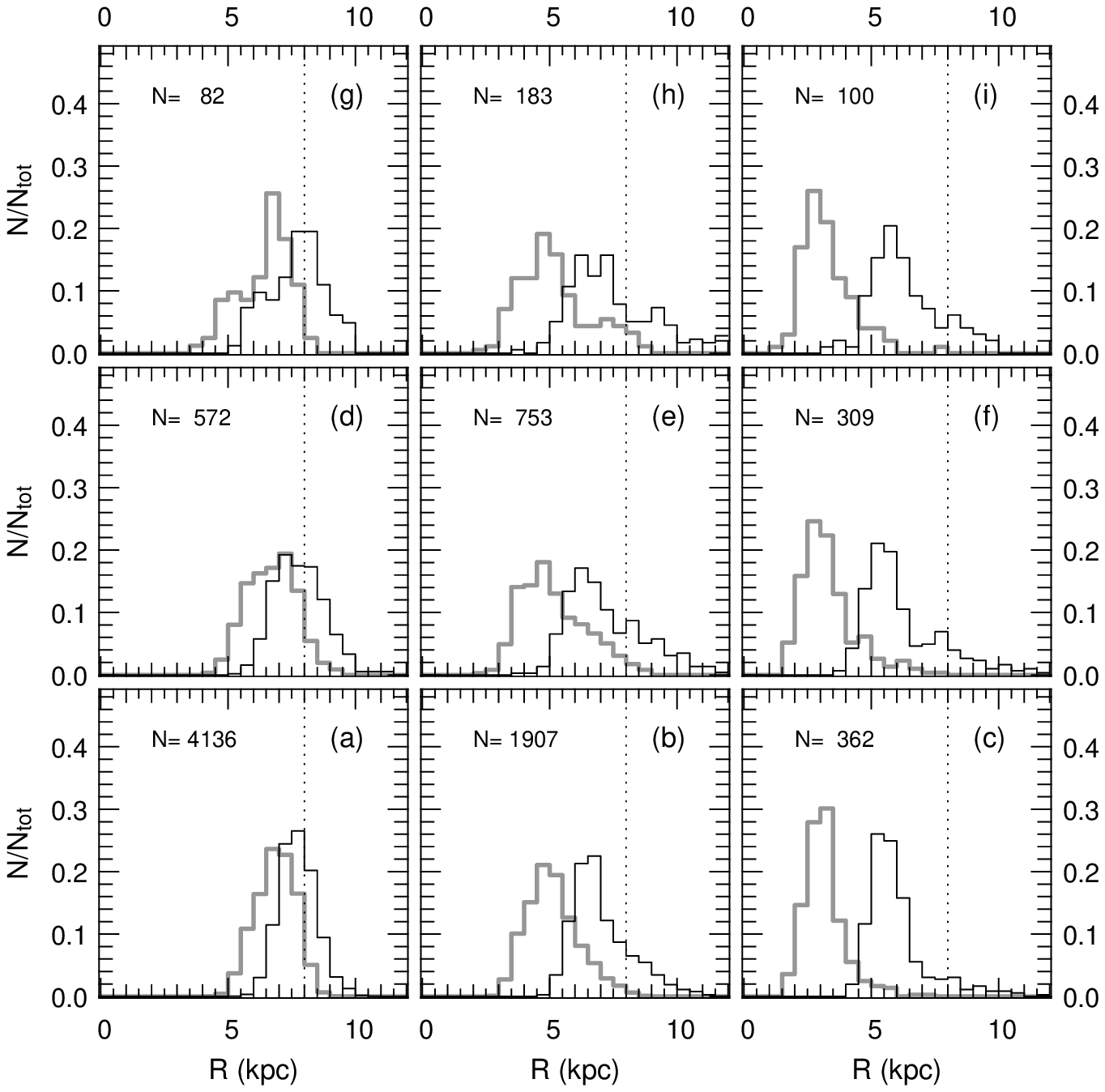}}
\centering 
\resizebox{\hsize}{!}
{\includegraphics[clip=,width=14cm]{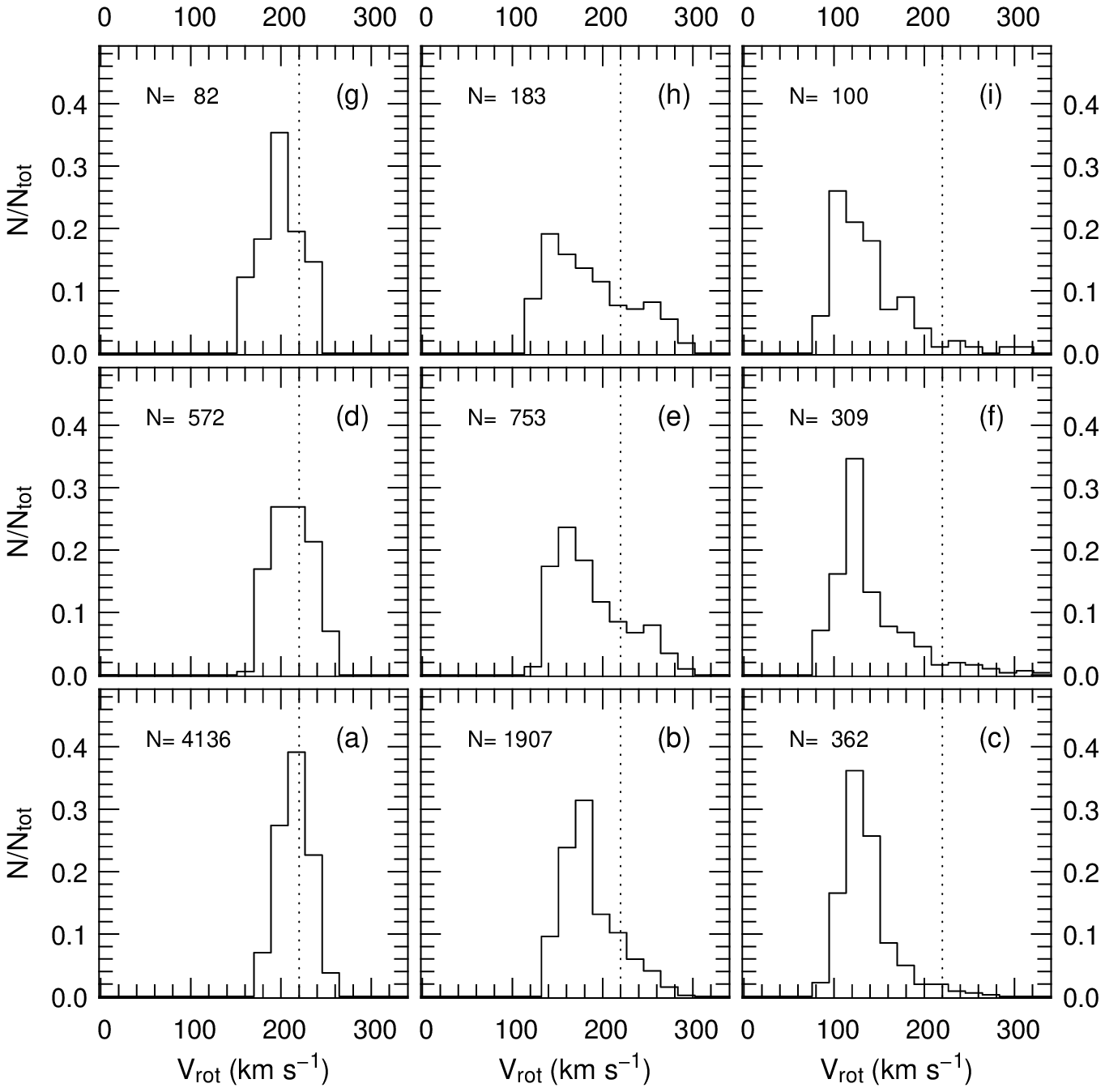}}
\caption{Upper panel: Perigalacticon (grey lines) and mean radius (black lines) 
distributions for the stellar samples defined by 
panels (a) through (i) in Figure~\ref{Zmax_e_sections}. The dotted line
indicates $R$=8~kpc. Lower panel
Rotational velocities $V_{\rm rot}$ for the stellar samples defined by 
panels (a)-(i) in Figure~\ref{Zmax_e_sections}. The dotted line indicates
$V_{\rm rot}$=220 km s$^{-1}$.
}
\label{Zmax_e_Vphi} 
\end{minipage}
\end{figure*}

By moving up from panel (a) to panel (g) the mean $R_p $ and
$R_m$ do not change significantly but their distributions
become broader (see Figure~\ref{Zmax_e_Vphi} top panel), whereas the $V_{\rm rot}$
distribution slightly decreases its mean from 220~km~s$^{-1}$ to
200~km~s$^{-1}$ (see Figure~\ref{Zmax_e_aFe} bottom panel).  
The Fe-distribution is found to shift to lower abundances
(with a peak around $-$0.6~dex, Figure~\ref{Zmax_e_aFe} top panel), 
and the average [$\alpha$/Fe] increases by
$\sim$0.1~dex.  All the above properties are indicative of a sample
dominated by local thick disc stars (panel g).  However, the presence of
kinematically heated old thin disc stars cannot be discarded.
Stars with high $Z_{\rm max}$ and low eccentricities must be on orbits 
which strongly oscillate through the Galactic plane, suggesting that they have
experienced some perturbations during their lifetimes. Feltzing \& Bensby 
\cite{feltzing} proposed the same interpretation for a subsample of 32 stars
which would lie in panel (d) of our Figure~\ref{Zmax_e_sections}. 
The cause for this might be identified in the kinematical heating mechanism
proposed by some authors (e.g.  Villalobos \& Helmi
\citealp{villalobos}, Bournaud et al.  \citealp{bournaud09},  
Minchev et al. \citealp{minchev12c} for different
processes) in order to explain the thickness of the disc.

\begin{figure}[t]
\centering
\resizebox{\hsize}{!}
{\includegraphics[clip=,width=9cm]{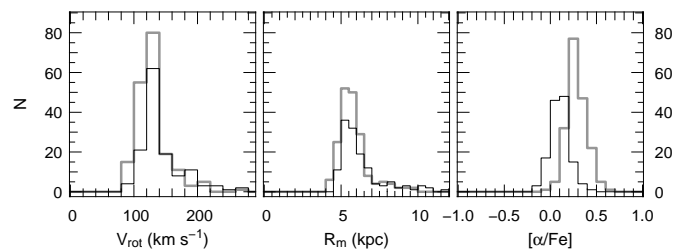}}
\caption{Distributions in $V_{\rm rot}$ and $R_m$ of the high
metallicity tail ([Fe/H$\ge-0.4$~dex, black line) and low metallicity tail
([Fe/H]$<-0.5$~dex, grey line) of the distribution in Figure~\ref{Zmax_e_aFe}
top, panel (c).}
\label{Vrot_R_aFe} 
\end{figure}

Focusing now on panel (c) of Figures~\ref{Zmax_e_aFe} and
\ref{Zmax_e_Vphi} (upper panels), we identify a population with high
eccentricities but confined close to the disc plane ($Z_{\rm
max}<1$ kpc, $0.4>e>0.6$).
Stars in this group have small perigalactic radii ($R_p \sim$3~kpc) and
small mean radii ($R_m\sim$6~kpc).  Due to their high eccentricities and the
large fraction of stars coming from the inner disc, the
observed V$_{\rm rot}$ distribution (Figure~\ref{Zmax_e_Vphi}, lower panel)
peaks at $\sim$120~km s$^{-1}$, and has a tail extending up to more than 200
km s$^{-1}$.  In addition, stars in this panel show a broad
Fe-distribution (Figure~\ref{Zmax_e_aFe}, upper panel) with a hint
of bimodality.  The latter differs
significantly from the metallicity distributions in panels (a) or (b) (a
Kolmogorov-Smirnov test showed that the probability that the distribution in
panel (c) is drawn from the populations of panel (a) or (b) is both lower
than $10^{-17}$). Note also that, by moving upward from
panel (c) to panel (i), the high [Fe/H] tail of the distribution
progressively disappear, leaving a Fe-distribution qualitatively
similar to the thick disc one. This means that in panel (c) there are most
probably more than one population, i.e.  stars that might belong to both
the thin and thick discs.  While stars with low Fe abundance
([Fe/H]$<-0.5$~dex) have all the characteristic to be identified as thick disc,
stars  with high Fe abundance ([Fe/H]$>-0.4$~dex) are likely to be thin-disc
stars scattered outward from the inner part of the Galaxy.
Being such stars drawn from panel (c) their kinematic is determined and 
thick-disc-like, but their chemical abundance is typical of the thin-disc.
Therefore, they have no clear thin- thick-disc classification.
This feature is highlighted in Figure~\ref{Vrot_R_aFe} where the distributions
in $V_{\rm rot}$, $R_m$ and [$\alpha$/Fe] of the 
two tails are shown separately.  Kinematically, the stars belonging to the
two tails show no significant differences, whereas they have distinct
chemical abundances in [Fe/H] and [$\alpha$/Fe].

Such stars having thick-disc kinematics and thin-disc chemical
abundances might have been kinematically heated and/or migrated 
by a mechanism which scatters out stars 
from the inner parts of the Galaxy, and could
be identified as the gravitational actions of the spiral arms  (Sch\"onrich \& Binney
\citealp{schoenrich}, Ro{\v s}kar et al. \citealp{roskar08}) and/or the
Galactic bar (Minchev \& Famaey
\citealp{mf10}, Minchev et al. \citealp{minchev11a}, Brunetti et al.
\citealp{brunetti11}).

The identification of two populations coming from the inner parts of the
Galaxy, lying in the Galactic plane, but having different metallicities,
suggests that thick-disc and dynamically heated and/or migrated stars are separate
populations with different origin and evolution. This observation is
consistent with the previous findings by Wilson et 
al. \cite{wilson}, Liu \& van de Ven \cite{liu12} and Kordopatis et al.
\cite{kordopatis1}, 
based on the eccentricity distribution of thick disc stars and
challenges scenarios in which the thick disc formed solely by the outward
migration of stars born in the inner disc (Sch\"onrich \& Binney,
\citealp{schoenrich}), while
on the other hand supports the predicted existence and action of such a mechanism
(see also Grenon \citealp{grenon} and Trevisan et al. \citealp{trevisan}). 
Our conclusion is also in agreement with the recent work by Minchev et al. 
\cite{minchev12c}, who suggested that both mergers at early phases and the 
effect of a central bar at later times are necessary to
explain the presence of stars with thick-disc chemistry and kinematics
currently found in the solar vicinity.

The fact that we find such stars in panel (c) does not exclude the
possibility that heated or migrated stars might be found in other regions of
the $e-Z_{\rm max}$ plane.  For instance, as mentioned in
Sec.~\ref{e-Zmax_sec} migrated stars can move away from their original
Galactic radius and conserve their eccentricity, making them
indistinguishable from the locally born stars. 
In such case only the chemical composition can reveal the difference between
migrated and local stars, but in our sample, the
identification of such stars is challenging, because to draw a clear
chemical signature requires a number of measured elements and a precision in
abundance higher than the ones provided by RAVE.\\
On the other hand, kinematically heated stars would have high
probability to get into high eccentric orbits. In doing so, they would
enrich the poorly populated tail of the eccentricity distribution of the
local stars, and broad the [Fe/H] distribution, making more likely to see a
bimodality within the error. This make panel (c) a more favourable 
region where to look for such stars.

\section{Discussion and conclusions}

We selected two samples of 2167 and 9131 giant stars from the RAVE chemical catalogue having good
quality spectra (S/N$>$75 and S/N$>$60, respectively called ``SN75" and
``SN60" sample) and known kinematics. We applied to the SN75 the
kinematic criteria by Gratton et al. \cite{gratton03} (G03) aiming at
disentangling different populations of the Milky Way. Following G03, we divided
the SN75 sample into 
thin disc, dissipative component and accretion component.
We verified first that our data (obtained from medium resolution spectra)
matches well the results by G03 obtained from high-resolution
spectra, confirming the reliability of the RAVE kinematic
and chemical abundances.

Our thin disc and dissipative (mostly thick disc) samples selected by a
pure kinematical criteria turned out to have similar distance and
eccentricity distributions as those reported by Lee et al. \citealp{lee11}
for a G-dwarf SEGUE sample where the thin and thick disc stars were selected
based on a pure chemical criterion.

We also notice that thin disc, dissipation and accretion components
partially overlap in several parameters.  Such overlaps are to be
expected if the processes currently debated in the literature (accretion, heating and
stellar migration) are at play during the evolution of the Milky Way. This
is particularly true for thin and thick disc (the latter is identified with
the dissipation component) which kinematically overlap one another in a way
that makes it difficult (if not impossible) to find selection criteria capable
of disentangling them.  This realization pushed us to drop the searching for
better selection criteria and led us to look for an alternative
approach.

The novelty of this work is the introduction of the analysis of the $e-Z_{\rm max}$ 
plane\footnote{A different use of the $e-Z_{\rm max}$ plane was previously
done by Feltzing \& Bensby \cite{feltzing}.}
which permits us to group stars having similar orbits. In fact, eccentricity
determines the shape of the orbit and $Z_{\rm max}$ determines the
oscillation amplitude of the stars in the Galactic plane, i.e., the
probability of the star to populate the region 
typically occupied by the thick disc.
We applied this analysis to the SN60 sample, and divided it in 9 
groups on the $e-Z_{\rm max}$ plane.
By studying the distributions of $R_m$, $R_p$,
$V_{\rm rot}$, [Fe/H] and [$\alpha$/Fe] of these
groups, we found stellar samples which identify nicely the thin and thick
discs, as well as a third sample lacking a clear thin- thick-disc
classification.

In particular, we identified an interesting subsample of stars with large
eccentricities ($0.4 < e < 0.6$), low $Z_{\rm max}$ (below 1~kpc), with
guiding radii in the inner disc ($R_p \sim$ 3 kpc and
$R_m \sim$ 6~kpc), and which shows broad distributions in [Fe/H] and
[$\alpha$/Fe].  The
[Fe/H]-poor tail of the distribution is composed of stars which have
properties similar to thick disc stars, whereas the high [Fe/H] tail 
shares with them the kinematical signature but differs clearly in Fe 
abundance as well as in $\alpha$ enhancement ([$\alpha$/Fe]).
A sample of stars with similar orbits and different metallicities 
suggests the existence of a heating/migration mechanism which
pushes stars from inner part of the Galaxy outward. However, we cannot
discard the possibility that such stars have been kinematically heated by
merging events or belong to a merged satellite themselves (see e.g. Helmi et
al. \citealp{helmi}).

Our results support a number of previous works, which
have shown that the thin and thick discs overlap kinematically. Even
if a clear separation did exist in the past, heating from spiral arms
(e.g., Carlberg \& Sellwood \citealt{carlberg85}, Minchev \& Quillen
\citealp{minchev06}), a central bar (e.g.,
Minchev \& Famaey \citealt{mf10}), giant molecular clouds (e.g., Jenkins \&
Binney \citealt{jenkins90}) would have  
blurred the kinematical borders of the two discs and finally merge 
them. Additionally, radial migration due to transient spiral density
waves (Sellwood \& Binney \citealp{sellwood02}), 
the interaction among multiple spirals
(Minchev \& Quillen \citealp{minchev06}) 
or bar and spirals (Minchev \& Famaey \citealp{mf10}, Minchev et al.
\citealp{minchev11a}, Brunetti et al. \citealp{brunetti11}), 
and the effect of minor mergers (Quillen et al. \citealp{quillen09},
Bird et al. \citealp{bird11}), introduces an even harder problem, since by changing their
angular momenta, stars arriving to the solar neighborhood have
kinematics indistinguishable from those born in-situ.
In case of satellite accretion (e.g., Villalobos \& Helmi
\citealt{villalobos}, Abadi et al. \citealp{abadi}) the kinematical
borders would also be blurred (because the consequent kinematical heating of
the pre-existing disc) with, in addition, a chemical overlap of the accreted stellar
populations to the Galactic population.
High-resolution spectroscopic surveys would be necessary to distinguish
the chemical fingerprints of the extra-galactic population from the one born
in situ.

Both stellar heating and migration are time-dependent processes.
Therefore, in the hypothesis that originally thin and thick disc 
were kinematically distinguishable, the natural question arises: 
How long does it take to
delete the original kinematical borders between them? 
On the basis of our kinematic and chemical data, we infer that
this time should be rather long (comparable to the Galaxy's lifetime), 
given that the RAVE sample contains 
stars with signs of kinematic diffusion together with the expected   
thin- and thick-disc populations.
Models of our Galaxy together with tools that
create synthetic surveys (Sharma et al., \citealp{sharma}) 
will be employed in a next work in order to compare observations 
with up-to-date models of the Milky Way.
Qualitative and quantitative
comparisons of our data with detailed
chemodynamical models which follow the evolution of a Milky Way-like 
disc for the entire expected thin- and thick-disc lifetimes, are
necessary to understand better the formation and evolution of the
Galactic discs.

\begin{acknowledgements}

C.B. thanks J. Binney, A. Just and B. Anguiano for their useful comments.
We acknowledge funding from Sonderforschungsbereich SFB 881 ``The Milky Way
System" (subproject A5) of the German Research Foundation (DFG).  Funding
for RAVE has been provided by: the Australian Astronomical Observatory; the
Leibniz-Institut fuer Astrophysik Potsdam (AIP); the Australian National
University; the Australian Research Council; the French National Research
Agency; the German Research Foundation (SPP 1177 and SFB 881); the European
Research Council (ERC-StG 240271 Galactica); the Istituto Nazionale di
Astrofisica at Padova; The Johns Hopkins University; the National Science
Foundation of the USA (AST-0908326); the W.  M.  Keck foundation; the
Macquarie University; the Netherlands Research School for Astronomy; the
Natural Sciences and Engineering Research Council of Canada; the Slovenian
Research Agency; the Swiss National Science Foundation; the Science \&
Technology Facilities Council of the UK; Opticon; Strasbourg Observatory;
and the Universities of Groningen, Heidelberg and Sydney.  The RAVE web site
is at http://www.rave-survey.org.

\end{acknowledgements}


\begin{thebibliography}{}

\bibitem[2003]{abadi} Abadi, M.~G., Navarro, 
J.~F., Steinmetz, M., \& Eke, V.~R.\ 2003, \apj, 597, 21 
%
\bibitem[2011]{adibekyan} Adibekyan, V. Zh., Santos, N. C., Sousa, S. G., Israelian,
G., 2011, A\&A, 535, 11
%
\bibitem[2003]{bensby03} Bensby, T., Feltzing, S., \& Lundstr{\"o}m,
I.\ 2003, \aap, 410, 527
%
\bibitem[2005]{bensby05} Bensby, T., Feltzing, S., Lundstr{\"o}m, I.,
\& Ilyin, I.\ 2005, \aap, 433, 185
%
\bibitem[2012]{bensby12} Bensby, T. and Feltzing, S., 2012, EPJ Web of Conferences, Vol. 19,
id.04001
%
\bibitem[2012]{bijaoui} Bijaoui, A., Recio-Blanco, A., 
\& de Laverny, P. and Ordenovic, C., 2012, Statistical methodology, 9, 55
%
\bibitem[2011]{bird11} Bird, J.~C., Kazantzidis, 
S., \& Weinberg, D.~H.\ 2011, \mnras, 2140
%
\bibitem[2011]{boeche11} Boeche, C., Siebert, A., Williams, M., et al., 2011, AJ, 142, 193
%
\bibitem[2011]{bovy11} Bovy, J., Rix, H.-W., Liu, 
C., et al.\ 2011, arXiv:1111.1724
%
\bibitem[2009]{bournaud09} Bournaud, F., Elmegree, B. G., Martig, M. 2009, \apj, 694, L158
%
\bibitem[2010]{breddels} Breddels, M.~A., Smith, M.~C., Helmi, A., et
al.\ 2010, \aap, 511, A90 
%
\bibitem[2012a]{brook12a} Brook, C.~B., Stinson, 
G., Gibson, B.~K., et al.\ 2012, \mnras, 419, 771 
%
\bibitem[2011]{brunetti11} Brunetti, M., Chiappini, C., Pfenniger, D. 2011, A\&A 534, 75
%
\bibitem[2011]{burnett} Burnett, B., Binney, J., Sharma, S., et al.\
2011, \aap, 532, A113 
%
\bibitem[1985]{carlberg85} Carlberg, R.~G., 
\& Sellwood, J.~A.\ 1985, \apj, 292, 79 
%
\bibitem[2011]{casagrande} Casagrande, L., Sch\"onrich, R., Asplund, M.,
et al., 2011, A\&A, 530, 138
%
\bibitem[2012]{cheng12} Chen, J. Y., Rockosi, C. M., Morrison, H. L. et al.\ 2012, \apj, 746, 149 
%
\bibitem[2009]{chiappini09} Chiappini, C., 2009, The Galaxy Disk in
Cosmological Context, Proceedings of the International Astronomical Union,
IAU Symposium, Volume 254. Edited by J. Andersen, J. Bland-Hawthorn, and B.
Nordstr\"om, p. 191-19
%
\bibitem[1998]{dehnen98} Dehnen, W., Binney, J., 1998, MNRAS, 294, 429
%
\bibitem[1998]{dehnen98b} Dehnen, W., \& Binney, J.~J.\ 1998,
\mnras, 298, 387 
%
\bibitem[2012]{dimatteo12} Di Matteo, P., Qu, 
Y., Lehnert, M.~D. \& van Driel, W.\ 2012, Assembling the Puzzle of the Milky Way, Le
Grand-Bornand, France, Edited by C.~Reyl{\'e}; A.~Robin; M.~Schultheis; EPJ
Web of Conferences, Volume 19, id.04002, 19, 4002 
%
\bibitem[2008]{feltzing} Feltzing, S., \& Bensby, T., Phys. Scr., 2008, T133, 014031
%
\bibitem[2011]{forbes} Forbes, J., Krumholz, 
M.~R., \& Burkert, A.\ 2011, arXiv:1112.1410 
%
\bibitem[1998]{fuhrmann98} Fuhrmann, K., 1998, A\&A, 338, 161
%
\bibitem[2008]{fuhrmann08} Fuhrmann, K., 2008, MNRAS, 384, 173
%
\bibitem[1990]{jenkins90} Jenkins, A., \& Binney, J.\ 1990, \mnras, 245, 305
%
\bibitem[1996]{gratton96} Gratton, R., Carretta, E., Matteucci, F., \&
Sneden, C., 1996, ASPC, 92, 307
%
\bibitem[2000]{gratton00} Gratton, R. G., Carretta, E., Matteucci, F., \&
Sneden, C., 2000, A\&A, 358, 671
%
\bibitem[2003]{gratton03} Gratton, R. G., Carretta, E., Desidera, S.,
Lucatello, S., Mazzei, P. and Barbieri, M., 2003, A\&A, 406, 131 
%
\bibitem[1999]{grenon} Grenon, M.\ 1999, Astrophys. Space Science, 265, 331
%
\bibitem[2011]{guedes11} Guedes, J., Callegari, 
S., Madau, P., \& Mayer, L.\ 2011, \apj, 742, 76 
%
\bibitem[2006]{helmi} Helmi, A., Navarro, 
J.~F., Nordstr{\"o}m, B., et al.\ 2006, \mnras, 365, 1309 
%
\bibitem[2012]{karatas12} Karata{\c s}, Y., Klement, R. J., 2012, NewA, 17, 22 
%
\bibitem[2011]{kordopatis} Kordopatis, G., Recio-Blanco, A., de
Laverny, P., et al.\ 2011, \aap, 535, A106 
%
\bibitem[2011b]{kordopatis1} Kordopatis, G., Recio-Blanco, A., de
Laverny, P., et al.\ 2011, \aap, 535, A107
%
\bibitem[2008a]{lee08a} Lee, Y. S., Beers, T. C., Sivarani,
T., 2008, AJ, 136, 2050
%
\bibitem[2008b]{lee08b} Lee, Y. S., Beers, T. C., Sivarani,
T., 2008, AJ, 136, 2022
%
\bibitem[2011]{lee11} Lee, Y. S., Beers, T. C., Deokkeun, A., et al., 2011,
AJ, 738, 187
%
\bibitem[2012]{liu12} Liu, C., \& van de Ven, G.\ 2012,
\mnras, 425, 2144 
%
\bibitem[2008]{marigo08} Marigo, P., Girardi, L., Bressan, A., Groenewegen, M.
A. T., Silva, L., Granato, G. L., 2008, A\&A, 482, 883
%
\bibitem[2012]{matijevic} Matijevi{\v c}, 
G., Zwitter, T., Bienaym{\'e}, O., et al.\ 2012, \apjs, 200, 14 


%
\bibitem[2006]{minchev06} Minchev, I., \& Quillen, A.~C.\ 2006, \mnras, 368, 623
%
\bibitem[2010]{mf10} Minchev, I., \& Famaey, B.\  2010, \apj, 722, 112 
%
\bibitem[2011a]{minchev11a} Minchev, I., Famaey, B., Combes, F., et al.\ 2011, \aap, 527, A147
%
\bibitem[2011b]{minchev11b} Minchev, I., Famaey, 
B., Quillen, A.~C., \& Dehnen, W.\ 2011, arXiv:1111.0195
%
\bibitem[2012a]{minchev12} Minchev, I., Famaey, B., Quillen, A.~C., et al.\ 2012, arXiv:1203.2621, In
press
%
\bibitem[2012b]{minchev12b} Minchev, I., Famaey, 
B., Quillen, A.~C., et al.\ 2012, arXiv:1205.6475
%
\bibitem[2012c]{minchev12c} Minchev, I., Chiappini, C., Martig, M. 2012, arXiv:1208.1506, submitted
%
\bibitem[2011]{navarro11} Navarro, J.~F., Abadi, 
M.~G., Venn, K.~A., Freeman, K.~C., 
\& Anguiano, B.\ 2011, \mnras, 412, 1203 
%
\bibitem[2010]{nissen} Nissen, P.~E., \& Schuster, W.~J.\
2010, \aap, 511, L10 
%
\bibitem[2004]{nordstrom04} Nordstr\"om, B., Mayor, M., Andersen, J., et al., 2004, A\&A 418, 989
%
\bibitem[2012]{pilkington12} Pilkington, K., Few, C.~G., Gibson, B.~K.,
et al.\ 2012, \aap, 540, A56 
%
\bibitem[2011]{piontek11} Piontek, F., \& Steinmetz, M.\
2011, \mnras, 410, 2625 
%
\bibitem[2009]{quillen09} Quillen, A.~C., 
Minchev, I., Bland-Hawthorn, J., \& Haywood, M.\ 2009, \mnras, 397, 1599 
%
\bibitem[1993]{quinn93} Quinn, P.~J., Hernquist, 
L., \& Fullagar, D.~P.\ 1993, \apj, 403, 74
%
\bibitem[2006]{recioblanco} Recio-Blanco, A., 
Bijaoui, A., \& de Laverny, P.\ 2006, \mnras, 370, 141 
%
\bibitem[2006]{reddy06} Reddy, B.~E., Lambert, 
D.~L., \& Allende Prieto, C.\ 2006, \mnras, 367, 1329 
%
\bibitem[2008]{roskar08} Ro{\v s}kar, R., 
Debattista, V.~P., Quinn, T.~R., Stinson, G.~S., 
\& Wadsley, J.\ 2008, \apjl, 684, L79
%
\bibitem[2009]{blazquez09} S{\'a}nchez-Bl{\'a}zquez, P., Courty, S., Gibson, B.~K., 
\& Brook, C.~B.\ 2009, \mnras, 398, 591
%
\bibitem[2011]{scannapieco11} Scannapieco, C., Wadepuhl, M., 
Parry, O.~H., et al.\ 2011, arXiv:1112.0315
%
\bibitem[2011]{schlesinger11} Schlesinger, K.~J., 
Johnson, J.~A., Rockosi, C.~M., et al.\ 2011, arXiv:1112.2214
%
\bibitem[1993]{schuster93} Schuster, W. J., Parrao, L., Contreras Martinez, M.
E., 1993, A\&AS, 97, 951
%
\bibitem[2009a]{schoenrich} Schoenrich, R., Binney, J., 2009a, MNRAS 396, 203
%
\bibitem[2009b]{schoenrich1} Sch{\"o}nrich, R., \& Binney, J.\
2009b, \mnras, 399, 1145 
%
\bibitem[2002]{sellwood02} Sellwood, J.~A., \& Binney, J.~J.\
2002, \mnras, 336, 785 
%
\bibitem[2011]{sharma} Sharma, S., 
Bland-Hawthorn, J., Johnston, K.~V., \& Binney, J.\ 2011, \apj, 730, 3 
%
\bibitem[2011]{siebert11} Siebert, A., Williams, M. E. K., Siviero, A., et al., 2011,
AJ, 141, 187
%
\bibitem[2006]{steinmetz} Steinmetz, M., Zwitter, T., Siebert, A., et al., 2006,   AJ 132, 1645
%
\bibitem[2012]{steinmetz12} Steinmetz, M., 2012, AN 5/6, 523 
%
\bibitem[2011]{stinson} Stinson, G., Brook, C., Prochaska, J.~X., et al.\ 2011, arXiv:1112.1698 
%
\bibitem[1995]{teuben} Teuben, P.J., The Stellar Dynamics Toolbox NEMO, in: 
Astronomical Data Analysis Software and Systems IV, 
ed. R. Shaw, H.E. Payne and J.J.E. Hayes. (1995),
PASP Conf Series 77, 398
%
\bibitem[2011]{wilson} Wilson, M.~L., Helmi, 
A., Morrison, H.~L., et al.\ 2011, \mnras, 413, 2235 
%
\bibitem[2011]{trevisan} Trevisan, M., Barbuy, B., Eriksson, K., et
al.\ 2011, \aap, 535, A42 
%
\bibitem[2008]{veltz08} Veltz, L., Bienaym{\'e}, O., Freeman, K. C., et al., 2008, A\&A, 480, 753
%
\bibitem[2008]{villalobos} Villalobos A., Helmi A., 2008, MNRAS, 391, 1806
%
\bibitem[2009]{yanny} Yanny, B., Rockosi, C., Newberg, H.~J., et al.\ 2009, \aj, 137, 4377 
%
\bibitem[2008]{zwitter08} Zwitter, T., Siebert, A., Munari, U., et al., 2008, AJ, 136, 421
%
\bibitem[2010]{zwitter10} Zwitter, T., Matijevi\v{c}, G., Breddels, M. A., et al., 2010,
A\&A, 522, 54
\end{thebibliography}
\end{document}